\documentstyle[12pt,epsfig]{article}

\newcommand{\beq}{\begin{equation}}
\newcommand{\eeq}{\end{equation}}
\newcommand{\bea}{\begin{eqnarray}}
\newcommand{\eea}{\end{eqnarray}}

\newcommand{\tto}{\!\to\!}

\newcommand{\gsim}{\lower.7ex\hbox{$
\;\stackrel{\textstyle>}{\sim}\;$}}
\newcommand{\lsim}{\lower.7ex\hbox{$
\;\stackrel{\textstyle<}{\sim}\;$}}

\newcommand{\bibit}[1]{\bibitem{#1} \marginpar{\vspace*{.4cm}~~\tiny[#1]}}

\newcommand{\aver}[1]{\langle #1\rangle}

\newcommand{\La}{\overline{\Lambda}}

\newcommand{\mhad}{\mu_{\rm hadr}}

\newcommand{\GeV}{\,\mbox{GeV}}
\newcommand{\MeV}{\,\mbox{MeV}}
\newcommand{\matel}[3]{\langle #1|#2|#3\rangle}

\newcommand{\msp}[1]{\mbox{\hspace*{#1mm}~}}

\renewcommand{\bibit}[1]{\bibitem{#1}}

\begin{document}
\thispagestyle{empty}
\vspace*{-10mm}

\begin{flushright}
Bicocca-FT-04-14\\
UND-HEP-04-BIG\hspace*{.08em}03\\
LNF-04/17(P) \\ 
hep-ph/0410080 \\
\end{flushright}
\vspace*{7mm}

\begin{center}
{\LARGE{\bf
On the photon energy moments and their \vspace*{4mm}\\
`bias' corrections in \boldmath $B\to X_s+\gamma$
}}
\vspace*{14mm}

{\tt 
{\small {\rm Contributed to \vspace*{-1mm}}}\\ 
Flavor Physics\,\,\&\,\,CP\,\,Violation 2004, 
October 5-12, 2004, Daegu, Korea\vspace*{-1mm}\\
{\small {\rm and to }}\vspace*{-1mm}\\
CKM  $\,V_{ub}\,/\, V_{cb}\,$ at BELLE, October 12-13, 2004, 
Nagoya, Japan
}
\vspace*{8mm} 

{\large{\bf D.~Benson$^{a}\!$,\, I.I.~Bigi$^{\,a,b}$\, and 
N.~Uraltsev$^{\,c\,*}$}} \\
\vspace{4mm}

$^a$ {\sl Department of Physics, University of Notre Dame du Lac}
\vspace*{-.8mm}\\
{\sl Notre Dame, IN 46556, USA}\vspace*{.5mm}\\
$^b$ {\sl INFN, LNF, Frascati, Italy}\\
$^c$ {\sl INFN, Sezione di Milano, Milan, Italy}

\vspace*{16mm}

{\bf Abstract}\vspace*{-.9mm}\\

\end{center}

\noindent
Photon energy moments in $B \tto X_s + \gamma$ 
and the impact of experimental cuts are analyzed, including
the biases exponential in the effective hardness missed in
the conventional OPE. We incorporate the perturbative corrections
fully implementing the Wilsonian momentum separation ab initio. 
This renders perturbative effects numerically suppressed while leaving
heavy quark parameters and the corresponding light-cone distribution
function well defined and preserving their physical properties. The
moments of the distribution function are given by the heavy quark
expectation values of which many have been extracted from the $B\tto
X_c \,\ell\nu$ decays. The quantitative estimates for the biases in
the heavy quark parameters determined from the photon moments show
they cannot be neglected for $E_{\rm cut}\!\gsim\! 1.85\GeV$, and grow out
of theory control for $E_{\rm cut}$ above $2.1\GeV$. Implications for
the moments in the $B\tto X_c \,\ell\nu$ decays at high cuts are
briefly addressed.

\setcounter{page}{0}
\vfill

~\hspace*{-12.5mm}\hrulefill \hspace*{-1.2mm} \\
\footnotesize{
\hspace*{-5mm}$^*$
On leave of absence from
Department of Physics, University of Notre Dame,
Notre Dame, IN 46556, USA \\ 
\hspace*{-9pt} and from 
St.\,Petersburg Nuclear Physics
Institute, Gatchina, St.\,Petersburg 188300, Russia}
\normalsize

\newpage


\section{Introduction and Basics}
\label{INTRO}

The decays of beauty hadrons -- recognized for a long time as an area 
allowing high sensitivity probes of
fundamental dynamics -- have entered now also the realm of precision 
physics on both the experimental and
theoretical side. Inclusive semileptonic decays of $B$ mesons allow the 
accurate extraction
of the CKM parameters $V_{cb}$ and
$V_{ub}$ through their integrated rates and of other basic heavy quark 
parameters through the moments of their spectra. These heavy quark 
parameters  include the heavy quark masses $m_b$ and 
$m_c$ which are `external' to QCD, and
$B$ expectation values of local heavy quark operators like the kinetic 
$\mu_{\pi}^2$ or chromomagnetic $\mu_G^2$.  
The latter are controlled by the
strong forces and thus can in principle be calculated within QCD once 
theoretical control has been established over
nonperturbative dynamics. The motivation for knowing the heavy quark 
parameters  
accurately goes beyond wanting
to extract $V_{cb}$ and $V_{ub}$: their values provide challenging tests 
for the available treatment of 
strong dynamics. There has been significant recent progress in
this direction both on the theory and experiment 
\cite{delphi,babarprl,cleonew} side, with BABAR establishing 
new benchmarks by combining the
high statistics and quality of their data with a robust theoretical 
treatment.

Detailed studies of the shape of the photon spectrum in $B\!\to\! 
X_s+\gamma$ transitions produce
complementary information: (i) They allow a systematically independent 
determination of $\mu_{\pi}^2$ and
maybe even higher order terms. (ii) They lead to a more direct and thus 
potentially more accurate extraction of
$m_b$ than in semileptonic distributions  which depend on a combination 
of $m_b$ and $m_c$. (iii) They
provide immediate access to the distribution function of the $b$ quark 
inside the $B$ meson, the knowledge
of which reduces the model dependence in extracting $V_{ub}$ from 
inclusive $B\tto X_u\,\ell\nu $ transitions.

Claims have recently appeared in the literature challenging  
the classical OPE results relating the moments of the inclusive
distributions to the universal local heavy quark expectation values in
the decaying hadrons. They assert that the
expectation values extracted from the $b\tto c\,\ell\nu$ decays cannot
be related to the similar moments of the light-cone distribution
function controlling the heavy-to-light decays like $b\tto s\!+\!\gamma$,
due to intervention of the perturbative effects. We find no
justification for such revisions, and contend that the perturbative
corrections to the moments in both types of the decays can be treated
in the standard OPE approach, see Ref.~\cite{jet} for a recent
analysis. The heavy quark parameters extracted from 
$B\tto X_c\,\ell\nu\,$ definitely can be used to constrain the 
light-cone distribution function.

In practice, when measuring the photon energy 
spectrum in $B\tto X_s + \gamma$  experimenters have to impose lower
cuts on the photon energy to suppress backgrounds. Those cuts are 
presently in the
range $1.8$ to $2.1\GeV$ with little realistic hope for lowering them in 
the near future. Yet, as emphasized recently
\cite{misuse},  such cuts degrade the treatment of 
inclusive $B$ decays in the short-distance expansion. 
High cuts give rise to a new mass scale parameter ${\cal 
Q}$ in the expansion, which we refer to as `hardness', along with
the original parameter $m_b$. Say, for the inclusive $B\tto 
X_s\!+\!\gamma$ decays we have
\beq
{\cal Q} \approx M_B - 2 E_{\rm cut} \; .
\label{4}
\eeq 
Some nonperturbative contributions are then described in inverse powers 
of ${\cal Q}$ rather than $m_b$.
For high $E_{\rm cut}$ ${\cal Q}$ becomes too small for a $1/{\cal Q}$ 
expansion to be reliable.

The usually employed {\it practical} version of 
the operator product expansion (OPE) 
\footnote{A
detailed discussion of the principal simplifications employed in what we
conventionally refer to as the `practical OPE', can be found, e.g.\
in the review
by M.~Shifman \cite{shifdual}, and references therein.} 
often manifests this deterioration in the
power corrections explicitly -- in particular for semileptonic decays -- 
through the Wilson coefficients
of the higher-dimensional operators: they are functions of $E_{\rm 
cut}/m_b$, which effectively transform
$1/m_b^k$ into $1/{\cal Q}^k$ terms. A similar reduction  of the
effective
hard scale is present also in the purely 
perturbative contributions and can be traced through  
a decrease in the effective normalization scale for  
$\alpha_s$, once the BLM corrections are  included.

The OPE-based theory usually leaves out, however, exponential terms
like $e^{-M/\mu_{\rm had}}$ with $\mu _{\rm had}$ denoting the typical
momentum scale of nonperturbative QCD, and $M$ stands for the generic
high mass scale used in the expansion. Those are insignificant as long
as $M$ is driven by the $b$ quark mass scale and exceeds a few
$\GeV$. Yet, as the presence of high cuts causes ${\cal Q}$ to
decrease towards $1\GeV$, 
the terms which are exponential in ${\cal Q}$ rather than in
$m_b$ could quickly become significant.

The simplicity of two-body kinematics for $b\tto s+\gamma$ prevents the 
cuts from affecting the Wilson coefficients of the higher-dimensional 
operators. They do not directly reveal, therefore deterioration of the
expansion parameter  with high cuts. 
The cut dependence will surface and manifest this once the 
perturbative corrections to these
power terms are incorporated, yet this still is seen as a future level
for theory. 
Let us note that these exponential effects are not connected with the
potential growth of the perturbative corrections to the Wilson
coefficients when the cuts are raised. We have found indications 
that in the actual $B$ decays such perturbative
corrections are insignificant, once a physical renormalization scheme
is used.

The pilot study of Ref.~\cite{misuse} has shown that using the 
conventional OPE expressions
which lack such exponential terms introduces substantial
biases into the values of $m_b$ and
$\mu_\pi^2$ extracted from truncated photon energy 
moments -- i.e.\ moments evaluated with a cut on the photon energy -- 
for $E_{\rm cut} \!\sim\! 2\GeV$. Furthermore,  applying 
the corresponding bias corrections, even estimated in a simplified
manner,  resulted in a surprisingly good agreement in the 
OPE description between $b\tto s+\gamma$ and $b \tto c+\ell\nu$, and 
thus removed an otherwise apparent inconsistency. 
Therefore, in the present paper we want to present a more careful
theoretical evaluation of the photon moments with cuts, even if it is 
somewhat model-dependent, and to put them into a form suitable
for a global analysis of various $B$ decay data.

It had been appreciated quite some time ago that the  usual application
of the  OPE to 
radiative $B$ decays must be incomplete,
since it predicts the nonperturbative component of the photon energy 
moments to be independent of $E_{\rm cut}$.
Ref.~\cite{bauer} tried to assess the potential
effect of the biases, in our terminology, by applying general
inequalities based only on the positivity of the distribution
function. Being very general, the inequalities however are saturated by
quite unphysical functions having a support consisting of
isolated points. Therefore they are too weak to be of interest in
practice. Furthermore, these inequalities do not generally hold in
the presence of the perturbative corrections. Hence this approach has
limited utility.

In Ref.~\cite{bauer} it was also stated that the inequalities are
violated once the cut is placed too high, and this was interpreted as
physical evidence for the OPE breaking down for high cuts. We do not
think such an interpretation can be correct. 
The discrepancy noted there has a purely algebraic nature.  It was
actually caused by the Darwin term in 
Eq.~(5.11) of Ref.~\cite{bauer} entering with the wrong
sign. Correcting it leads to the inequality being trivially satisfied
at arbitrary $E_{\rm cut}$ and thus eliminates the contradiction, as
it has to be for a purely arithmetic bound.\footnote{The bound 
follows from the positivity of the integrals of
$F(z)\,(1\!-\!z)(1\!-\!tz)^2$ for arbitrary $t$, which is a quadratic
polynomial in $t$. Its determinant is therefore negative or zero,
which yields the inequality. Since the difference $1\!-\!P_n(E_{\rm
cut})$, with $P_n$ the weight used in Ref.~\cite{bauer} belongs to the
above class of functions ($t\!=\!1$), Eq.~(5.11) is trivially
satisfied.}

Alternative numerical estimates based on the AC$^2$M$^2$ ansatz 
\cite{ACM} led to a gross
underestimate: the biases emerged as negligible at cuts below $2\GeV$. The
perturbative corrections were not incorporated  at this
point, and too low nonperturbative parameters were used as a result of
following the pole scheme of HQET.

Similar complications plagued attempts to provide definite numerical
estimates in Ref.~\cite{kn}. The analysis there did not use
consistently defined Wilsonian heavy quark parameters, and operated in
terms of somewhat indefinite HQET analogies. The perturbative corrections in
that scheme are not stable and significantly change subtle effects
like biases in the photon moments. On the other hand, the values of
the nonperturbative parameters determining the assumed distribution
function were varied in an overly wide range including unphysically 
small values.
Therefore it was difficult to arrive at conclusive estimates, and
Ref.~\cite{kn} came up with no certain result: the estimates ranged
from the biases being totally negligible even at $E_{\rm cut}\!=\!2\GeV$ to
very significant down to $E_{\rm cut}\!=\!1.8\GeV$.

As described in Ref.~\cite{misuse}, relying on the Wilsonian
implementation of the OPE allows to derive more robust
statements. The perturbative corrections to physical effects are
moderate, and the nonperturbative parameters have definite values,
often known with
small numerical uncertainties. For instance, BaBar has recently
presented the results of a comprehensive fit to the moments they have 
measured in inclusive semileptonic $B$ decays, obtaining \cite{babarprl}:
\bea
\nonumber
m_b(1\GeV) &=& 4.61\pm 0.05|_{\rm exp} \pm 0.04|_{\rm HQE} 
\pm 0.02|_{\rm th} \GeV \\
\mu_\pi^2 (1\GeV) &=&0.447\pm 0.035|_{\rm exp} \pm 0.038|_{\rm HQE} \pm 
0.010|_{\rm th} \GeV^2 \; , 
\label{12}
\eea
even without incorporating independent theoretical constraints on the 
heavy quark parameters (see, e.g., \cite{bounds}).
The results are in
good agreement with earlier preliminary DELPHI values
\cite{delphi,delphiconf}.

The analysis of $B\tto X_s\!+\!\gamma$ is complicated
by the fact that it is induced not only by the genuinely local weak 
operator
\beq
O_7= \frac{e}{32\pi^2} m_b\,\bar{s}\sigma_{\mu \nu} F^{\mu 
\nu}(1\!+\!\gamma_5)b,
\label{14}
\eeq
but depends also on additional terms in the effective Lagrangian
\beq
L_{bsg}= \frac{4G_F}{\sqrt{2}}V_{tb}V^*_{ts} \sum_{i=1}^{8} c_i O_i
\label{16}
\eeq
(we use the standard definition of $O_i$ which is given, e.g., in
Ref.~\cite{kn}). In particular, decays mediated by the operator
$O_2\!=\!\frac{1}{4}\bar{s}^k \gamma_\mu(1\!-\!\gamma_5)b^l\,
\bar{c}^l\gamma^\mu(1\!-\!\gamma_5)c^k$  (with the superscripts $k$, 
$l$ denoting color indices
to be summed over)
contribute through a (real or virtual) $c\bar{c}$-pair converting into a 
photon.
The effect of operators other than $O_7$, most
notably of $O_2$, is essential for a precise evaluation of the
decay rate \cite{misiak}. However, it is expected that the normalized
photon energy moments are not changed noticeably. The
corresponding refinements of Refs.~\cite{llmw,kn} modify them by negligible
amounts and therefore are irrelevant in practice. While illustrating
their effect for completeness, we need not dwell on the theoretical
subtleties associated with their evaluation.

The effects not captured directly by the standard OPE also affect the
$b\tto c$ inclusive semileptonic decay rates, once the applied cut on
the charged lepton energy becomes too high thus decreasing the hardness 
${\cal Q}$.
These effects are conceptually similar to the biases in
the photon energy moments, however the size of the two cannot be
related to each other unambiguously.

The remainder of the paper will be organized as follows: 
Sect.~\ref{STANDARD} presents the standard OPE for the photon 
energy spectra and their moments, based on the Wilsonian
implementation with an explicit scale separation.  
We address in detail the nontrivial impact of experimental cuts on
the OPE in Sect.~\ref{BIAS} and how they induce biases in the
values of such truncated moments; the interplay with the perturbative
corrections is studied. A step-by-step illustration of the
computational procedure is provided. 
We discuss the effect of the non-valence 
four-quark operators on the inclusive $B\tto X_s+\gamma$
characteristics in Sect.~4. In
Sect.~\ref{SLBIAS} we comment on the corresponding bias 
effects in semileptonic
decays before summarizing in Sect.~\ref{SUMMARY}. While the numbers and
plots in the paper use the exact formulae, only simplified
expressions are presented in the main text. Many of the full 
expressions are too cumbersome and are relegated to the Appendices. In
\ref{WILSPECT} we describe in some technical detail the calculation of 
the photon spectrum in a formulation where the Wilsonian prescription 
for the OPE has been implemented from the start.  Additional non-$O_7$
perturbative corrections are given in A.2, and in A.3 we give
the numerical evaluation of the bias effects. 

While the present paper was finalized the preprints \cite{neubbsg} 
appeared which
analyzed the $B\tto X_s\!+\!\gamma$ decay heavily 
relying on a SCET-motivated
approach. We largely disagree with the findings of the author, which
basically claim large perturbative uncertainty in the decays and
challenge accurate calculability of the integrated
characteristics derived from the $B\tto X_s\!+\!\gamma$ spectrum. We have
added Sect.~2.1 to briefly address this controversy, however the detailed
discussion of the SCET-based approach goes beyond the scope of the
present paper.

\section{Standard OPE for the photon energy moments}
\label{STANDARD}

The purely perturbative spectrum for $b\!\to\!s+\gamma$ has been
calculated in Ref.~\cite{llmw} (in the pole scheme) 
to first order in $\alpha_s$ and
including the BLM corrections $\propto \!\beta_0\alpha_s^2\;$ 
($\,\beta_0\!=\!11\!-\! \frac{2}{3}n_f\!=\!9$). The results 
for the moments read
\bea
\nonumber
\aver{E_{\gamma}}^{\rm pert}|_{E_{\gamma}\!>\!E_{\rm cut}}
\msp{-5}&=&\msp{-5}
\frac{m_b}{2}\left[1-\left[a_1(x;\mu)+
\hat a_1(x)\right] \frac{\alpha_s}{\pi}
-\left[b_1(x;\mu)+ \hat b_1(x)\right]
\frac{\beta_0}{2}\left(\frac{\alpha_s}{\pi}\right)^2 \right]
\\
\aver{E_{\gamma}^2\!-\!\aver{E_{\gamma}}^2
}^{\rm pert}|_{E_{\gamma}\!>\!E_{\rm cut}} \msp{-5}&=&\msp{-5}
\left(\!\frac{m_b}{2}\!\right)^{\!2}\left[
\left[a_2(x;\mu)+
\hat a_2(x)\right]
\frac{\alpha_s}{\pi}+
\left[b_2(x;\mu)+ \hat b_2(x)\right]
\frac{\beta_0}{2}
\left(\frac{\alpha_s}{\pi}\right)^2 \right] \msp{5}
\label{21}
\eea
where 
\beq
x =\frac{2 E_{\rm cut}}{m_b}\;.
\label{24}
\eeq
Existing calculations of the perturbative corrections do not allow to
specify which $b$ quark mass enters $x$ in Eq.~(\ref{24}), since the 
results in
different normalization schemes differ only in order $\alpha_s^2$ 
(without $\beta_0$).
However, on general grounds
we know that the pole mass cannot appear in short-distance
calculations; additional arguments can be found in
Ref.~\cite{uses}. Therefore, we assume $m_b$ in Eq.~(\ref{24}) to be
the running $b$ quark mass normalized at the scale $\mu \sim 1\GeV$.

Terms proportional to $\hat a_1(x)$, $\hat b_1(x)$, $\hat a_2(x)$,
$\hat b_2(x)$ correspond to contributions from operators other than
the dominant $O_7$.
While those
are noticeable for the total decay width, they are expected to be 
insignificant for the photon energy spectrum.
For instance, they are found to shift the first and second 
moment by
about $-0.004\GeV$ and $0.0008\GeV^2$, respectively, which translates
into $\delta m_b\!\simeq\! -8\MeV$ and 
$\delta\mu_\pi^2\!\simeq\! 0.01\GeV^2$, i.e.\
far below the accuracy of the perturbative piece alone. 
Thus there is no 
need for a detailed discussion of these effects.

The values of the coefficients $a_1(x; \mu)$ -- $b_2(x; \mu)$ in
Eqs.~(\ref{21}) depend on the normalization point $\mu$. 
They can be fixed by the fact that
the total moments (including nonperturbative contributions) are independent
of $\mu$ and by the condition that, to a given
order in $\alpha_s$, the expressions at $\mu\!=\!0$ reproduce the
result in the pole scheme which assumes no infrared cutoff in the
perturbative contributions. This yields
\bea
\nonumber
a_1(x; \mu)\msp{-4}&=&\msp{-4} a_1(x;0)-
\frac{4}{3} C_F \frac{\mu}{m_b}- \frac{13}{18}  C_F 
\frac{\mu^3}{m_b^3}\\
\nonumber
b_1(x; \mu)\msp{-4}&=&\msp{-4} 
b_1(x;0) - \frac{4}{3} C_F \left(\ln{\frac{M}{2\mu}}+\,
\frac{8}{3}\,\right) \frac{\mu}{m_b}-
\frac{13}{18} C_F \left(\ln{\frac{M}{2\mu}}+\frac{319}{130}\right)  
\frac{\mu^3}{m_b^3}\\
\nonumber
a_2(x; \mu)\msp{-4}&=&\msp{-4}
a_2(x;0)-\frac{1}{3}C_F\frac{\mu^2}{m_b^2}
+\frac{4}{9}C_F\frac{\mu^3}{m_b^3} \\
b_2(x; \mu)\msp{-4}&=&\msp{-4}
b_2(x;0)-\frac{1}{3} C_F \left(\ln{\frac{M}{2\mu}}+
\frac{13}{6}\right) \frac{\mu^2}{m_b^2}
+\frac{4}{9} C_F
\left(\ln{\frac{M}{2\mu}}+ 2-\frac{1}{20}\right) \frac{\mu^3}{m_b^3}
\; .
\label{30}
\eea
($M$ is the normalization scale for $\alpha_s$ in the $\overline{{\rm MS}}$
 scheme; we take it to be $m_b$.) 
The analytic expressions for $a_{1,2}(x;0)$ and
$b_{1,2}(x;0)$ are given in Appendix \ref{WILSPECT},
Eqs.~(\ref{269}). 

The expressions in Eqs.~(\ref{30}) are approximate in their
$\mu$-dependence -- even to a given order in $\alpha_s$ -- in two
respects.  They account for terms only through order $\mu^3/m_b^3\,$.
Since $\mu/m_b$ is a truly small parameter at $\mu\!\lsim\!  1\GeV$,
these are tiny corrections. Secondly and more importantly,
Eq.~(\ref{30}) is valid only at sufficiently low $E_{\rm
cut}$. Otherwise, the simple prescription described above does not
reproduce the actual $\mu$-dependence of the moments: it introduces
systematic shifts that are the perturbative analogue of the biases. In
the one-loop approximation (including all its BLM corrections) this
happens for $2E_{\rm cut}\!>\!m_b\!-\!2\mu$, i.e.\ where the hardness
${\cal Q}$ descends below $2\mu$. At high $E_{\rm cut}$ these
`perturbative biases' become significant.

Therefore we do not rely on the simplified expressions (\ref{30}), but
rather compute the moments explicitly integrating the perturbative
spectrum calculated with the Wilsonian cutoff at the separation scale
$\mu$. We have derived it both to order $\alpha_s$ and to all orders
in BLM, as described in Appendix \ref{WILSPECT}. The analytic
expressions for the first-order moments $a_i(x;\mu)$ are reasonably
compact. For the BLM corrections the analytic expressions have also
been obtained, yet they are too lengthy and we do not give them
explicitly in this paper. We have checked that at low cuts the
$\mu$-dependence indeed follows Eqs.~(\ref{30}).

The Wilsonian spectrum and, in particular its variation upon changing
$\mu$ has some remarkable features. In contrast to the usual spectrum,
it contributes even above  $m_b(\mu)/2$. The center of gravity of 
the spectrum upon varying $\mu$ does {\sf not} change to the first 
approximation; modifications arise in order 
$\delta \mu^2/m_b$, as mandated by the OPE applied to perturbative
effects.  \vspace*{1.5mm}

The nonperturbative pieces in the energy moments are given by
\bea
\langle E_{\gamma} \rangle^{\rm power}
\msp{-3}&=&\msp{-3}
\frac{\mu_{\pi}^2-\mu_{G}^2}{4m_b}-
\frac{5\rho_D^3-7\rho_{LS}^3}{12m_b^2}-
\frac{c_2}{c_7}\,\frac{\rho_{LS}^3}{54 m_c^2 }
\label{40}\\
\langle E_{\gamma}^2\!-\!\langle E_{\gamma} \rangle^2
\rangle^{\rm power} \msp{-3}&=&\msp{-3}
\frac{\mu_{\pi}^2}{12}
-\frac{2\rho_D^3-\rho_{LS}^3}{12m_b}+
{\cal O}\left(\frac{1}{m_b^2}\right) \; .
\label{41}
\eea
We have calculated them for the local weak decay operator
$O_7$; our results agree with the expressions in the revised version of
Ref.~\cite{bauer}. In principle, there are perturbative corrections to the
coefficients of these power terms; they have not been calculated for
realistic decays. 

The last term in Eq.~(\ref{40}) stands for certain effects of the
nonperturbative conversion of a $\bar cc$ pair into a photon where the
underlying weak decay is $b\!\to\!c\, \bar{c}s$. It scales only like
$1/m_c^2$. Such terms cannot arise in totally inclusive rates; they
rather represent final state interaction among the channels
mediated by $O_2$ and $O_7$.

One should note that  there exists no (local) OPE for such rescattering
widths. Thus, for instance, there is no reason for
$m_c$ entering there to be a short-distance charm mass. One may think
that the meson mass $M_D$ is actually appropriate there; this choice is
adopted in our numerical estimates.  Fortunately,
again this effect is negligible, yielding only
$\delta\aver{E_\gamma}\simeq -0.003\GeV$, i.e.\ potentially shifting
$m_b$ by as little as $6\MeV$.

Without accounting for the perturbative corrections to the
Wilson coefficients
of higher-dimen\-sional (nonperturbative) operators the total moments are
given by the sum of the perturbative and nonperturbative
contributions. In the absence of the former, the energy moments
obtained in the $1/m_b$ expansion do not depend on $E_{\rm cut}$,
which clearly cannot hold for the actual spectrum. Therefore, the 
thus computed moments are biased:
\bea
\nonumber
\aver{ E_{\gamma}}\msp{-2}& \simeq &\msp{-2} 
\aver{ E_{\gamma}}^{\rm pert} + \aver{ E_{\gamma}}^{\rm
power} +\mbox{$\frac{1}{2}$}\tilde\delta m_b, \\ 
\aver{E_{\gamma}^2\!-\!\aver{E_{\gamma}}^2}\msp{-4}& \simeq &\msp{-4} 
\aver{E_{\gamma}^2\!-\!\aver{E_{\gamma}}^2}^{\rm pert}\!+\!
\aver{E_{\gamma}^2\!-\!\aver{E_{\gamma}}^2}^{\rm
power}\!-\mbox{$\frac{1}{12}$}\tilde\delta \mu_\pi^2.
\label{48}
\eea
Note that the biases $\tilde\delta m_b$,  $\tilde\delta \mu_\pi^2$ are
defined as the corresponding corrections to the moments calculated in
the standard OPE in terms of the actual heavy quark parameters. We 
have introduced them with the coefficients motivated by the 
relation between the moments and the heavy quark parameters in the
large-$m_b$ limit. For this reason the biases are close -- yet strictly
speaking not identical -- to the shifts to be
applied to the values of $m_b$ and $\mu_{\pi}^2$ naively extracted
using the conventional prescription. In Sect.~3 we formulate the
suggestions  for evaluating these biases.

\subsection{Comments on the literature}

When the present study was completed  papers \cite{neubbsg} appeared
which addressed inclusive $b\tto s+\gamma$ decays from a different
perspective motivated by the technique  of the so-called
`Soft-collinear effective theory' (SCET).\footnote{There are actually
a number of competing versions of SCET, sometimes marked with the
numbers: SCET-I, SCET-II, etc. A partial list of references can
be found, for instance, in Refs.~\cite{neubbsg}.}
\,Their basic
statement appears rather paradoxical -- the perturbative
uncertainties were claimed to be quite significant, in conflict with our
findings. 
Since we have performed explicit  calculations,
we conclude the large
uncertainties in the treatment of Refs.~\cite{neubbsg} are
mainly an artefact of their approach, intrinsically associated with the
adopted SCET framework. 
A critical discussion of the latter goes beyond
the scope of our paper. Instead, we shall try to elucidate the problem
for the case at hand with minimal technical details. 
With SCET being a fast evolving field aiming at developing
a uniform approach to the light-front processes in $B$ decays ($b\tto
u\,\ell\nu$, $b\tto s\!+\!\gamma$), certain controversial issues
remain there. In particular, we may not agree with some
statements in SCET. It is therefore difficult to provide an
unbiased thorough review of the literature on the subject. As a
partial solution, we refer the reader to the recent papers
\cite{mannel1,beneke2,neubbsg} for the novel applications and for
references to the earlier papers and to the alternative approaches.

The so-called Soft-Collinear Factorization in QCD is 
highly emphasized in Refs.~\cite{neubbsg} as a theoretical advantage;
according to it the spectrum is represented in the form involving
the `hard' function $H$, jet function $J$ and and the `soft' `shape
function' $S$:
\beq
\frac{{\rm d}\Gamma}{{\rm d} E_\gamma} \propto H\cdot J\otimes S\,, 
\label{r210}
\eeq
where $\otimes$ symbolically denotes a convolution. $S$ is
just the familiar nonperturbative heavy quark distribution function; in
our Wilsonian scheme it is $F(k_+)$. In fact, the usual factorization
of the `soft' modes employed in the Wilsonian OPE (i.e., between
the effects from small and large distances) is sufficient for this
purpose. 
Moreover, $ H\cdot J$ is nothing but the (Wilsonian) perturbative
spectrum $\frac{{\rm d}\Gamma^{{\rm pert}}}{{\rm d} E_\gamma}$.
Contrary to the impression given in
Refs.~\cite{neubbsg}, an expression for the total spectrum in this form
was presented already in 1993 \cite{motion} (see also Ref.~\cite{bsg}), not
to mention a number of later papers.

Soft-collinear factorization, which goes back
to the Low theorem \cite{low} and its generalization in the Gribov-Low
theorem \cite{gribovlow}, in this context refers merely to how
to calculate
the perturbative spectrum itself. Regardless of anything else, it is
well known how to calculate the perturbative corrections in $B$
decays, and this {\sf has been} done.

The principal point of the analysis of Refs.~\cite{neubbsg}
is dividing
the dynamics driving inclusive $b\tto s\!+\!\gamma$ decays into three
domains of essentially different mass scale, namely $\sim \!\!m_b$,
$\sim \!\!\sqrt{m_b\mhad}$ and $\sim \!\mhad$. The perturbative
effects residing in the first two regions were treated separately
applying the  renormalization group methods which
implies a strong hierarchy among the scales.

A critical discussion of the validity of a similar approach to $b\tto
s\!+\!\gamma$
has already been given in Ref.~\cite{uses}. Rather than recapitulating
the details here, we mention only a few obvious points. With the genuine
Wilsonian cutoff at $1\GeV$ the gluon energy in the perturbative
diagrams cannot be below $1\GeV$. Kinematically, on the other hand, no
gluon with $E\!>\!\frac{m_b}{2}\!\simeq \!2.3\GeV$ can be emitted, and
even those with  an energy approaching $2\GeV$ are strongly
suppressed. This leaves a  limited phase space even for single
gluon emission and suppresses multiple emissions even more. The
approximation where both the individual gluon momenta and their
aggregate energy are much
smaller than the jet energy $m_b/2$, which is assumed in the  
Sudakov-type
treatment, can never be realized in $B$ decays even remotely.

The Sudakov treatment aims at $\log$-enhanced physical effects, it
allows their resummation. For that 
one employs a
reorganization of the perturbative series in powers of $1/\ln{Q}$
rather than $\alpha_s$, with $Q$ denoting a supposedly large
ratio of relevant scales. With $\alpha_s\!\cdot\! \ln{Q}$ then being
of order unity, the expansion in its powers has to be resummed.
Yet in $B$ mesons the `large' $\log$s hardly reach unity.
One cannot rely on an expansion valid at large $\ln{Q}$, when instead
$\,\ln{Q}\!<\!1$. The large-$\log$ series might not even converge in
that domain; there is no reason to expect such a procedure to yield
a meaningful result. This general problem becomes quite manifest in
$b\tto s\!+\!\gamma$ decays \cite{uses}:  for instance, the NLO
terms dominate over the leading order turning the physically
expected Sudakov suppression into an enhancement.

Furthermore, the renormalization group treatment per se is inefficient in
fixing the precise scale at which to evaluate the running
strong coupling, especially in the case of a limited momentum interval. 
 The concept of
the renormalization invariance itself has
nothing to say about whether, for instance, $\alpha_s(\Delta)$ appearing
in Refs.~\cite{neubbsg} is actually $\alpha_s(\Delta)$, or
$\alpha_s(2\Delta)$ or $\alpha_s(\Delta/2)$.
On the other hand,
$\alpha_s$ runs fast in the considered low-energy domain. This is a
major factor contributing to the large uncertainties of
Refs.~\cite{neubbsg}.
This ambiguity would only be resolved in the RG framework by
calculating the matching coefficients to higher orders.
 At the same time, the improvement achieved directly in 
QCD by adding straightforward BLM corrections, without subdividing 
the perturbative domain, does this far more reliably \cite{uses}.

Let us recall that a detailed illustration of how a formally correct
large-$\log$ renormali\-zation-group treatment yields a numerically
misleading result for an energy scale gap typical in $B$ physics, was
given already 10 years ago
\cite{comment}, in a similar context of the zero-recoil $b\tto c$
transitions. That time a similar NLO renorm-group improvement
accomplished by Neubert was claimed to yield a precision perturbative
factor $\eta_A\!=\!0.986\!\pm\! 0.006$
\cite{neubpert}, while  a thoughtful application of the direct
one-loop QCD calculation yielded $\eta_A\!\simeq\!0.96$ -- a value
later confirmed by the explicit higher-order calculations. 
We think the observed large uncertainties in the calculation
of the photon spectrum in Refs.~\cite{neubbsg} root in a 
similar misconception.

An additional advantage of our approach over SCET is that the (quite
significant) $1/m_b$ corrections, including those in the perturbative
coefficients, are automatically included; the KLN-type relations are
respected at every order, which does not necessarily hold in
the truncated RG approach. We emphasize that, using a commensurate
language, the required anomalous dimensions, the BLM corrections
to the `SCET matching coefficients' are all automatically incorporated
in the full QCD perturbative corrections we use; they actually can be
read off from the analytic expressions of Appendix A1.

The most accurate evaluation for $B$ decays is 
obtained in the full-QCD fixed-order perturbative
calculations. Implementing the Wilsonian approach significantly reduces
the corrections and makes them well behaved; the effect of the
running of the strong coupling in different kinematic domains 
is readily incorporated  
by calculating the BLM corrections in the same framework. This  
strategy had already been advocated in 
Ref.~\cite{uses} and is performed in this paper.

\section{Bias corrections}
\label{BIAS}

\subsection{Basic ansatz}

As described in Ref.~\cite{misuse} and stated in the Introduction,
there is no unambiguous way to determine precisely the biases in the photon
moments, unless the  
nonperturbative dynamics for $B$ mesons have been brought 
under full theoretical
control. On the other hand with $B$ mesons constituting the ground state 
we can expect the $b$ quark distribution function $F(k_+)$ 
describing its intrinsic
motion, to be essentially a positive and 
smooth function with a pronounced peak near  
$E_\gamma\!=\!m_b/2$. There is 
no unique choice for such a
function. Yet once the average and the width of the distribution
are fixed 
by the values of the heavy quark parameters, the function is sufficiently
constrained to allow reasonably accurate estimates of
the cut-induced biases for naturally shaped distributions.

For proper perspective let us recall a few basic facts. The 
tail of $F(k_+)$ corresponding to the very low
$E_\gamma$ and its behavior near maximal $k_+$ corresponding to
$E_\gamma$ approaching $M_B/2$ depend on the
high-dimension expectation values and are determined by the subtle
details of the bound-state dynamics. That  
implies that nothing reliable can be said about
biases at $E_{\rm cut}\!\gsim\! m_b/2$,  nor are they under full
theoretical control at very low 
$E_{\rm cut}$. Yet for low $E_{\rm cut}$ 
their size is sufficiently small to render them 
insignificant. Our prescription is therefore meant to be applied for 
$E_{\rm cut}$ in the range
from about $1.7\GeV$ to $2.1\GeV$. Those numbers follow from the 
expected values $m_b\!\simeq\!4.6\GeV$ and 
$\mu_\pi^2 \!\simeq\! 0.45\GeV^2$ \cite{babarprl,delphi}.

The relevant parameters describing the shape of the $b$ quark 
distribution function are its 
mean  and variance which (in terms of $E_\gamma$)  are  $m_b/2$
and $\mu_\pi^2/12$, respectively, in the heavy quark limit.
For finite quark mass they have  power corrections, which can be read 
off from Eqs.~(\ref{40})--(\ref{41}). For use  in $F(k_+)$ we can
then replace $m_b$ and $\mu_{\pi}^2\,$ by $\,\widehat{m_b}\,$ 
and $\,\widehat{\mu_{\pi}^2}\:$ with
\bea
\nonumber
\widehat{m_b} \msp{-4}& \simeq &\msp{-4} m_b +
\frac{\mu_{\pi}^2-\mu_{G}^2}{2m_b}-
\frac{5\rho_D^3-7\rho_{LS}^3}{6m_b^2}\\
\widehat{\mu_{\pi}^2}\msp{-4}& \simeq &\msp{-4}
\mu_{\pi}^2-r_3\frac{2\rho_D^3-\rho_{LS}^3}{m_b}\;.
\qquad \label{50}
\eea
In the strict $1/m_b$ expansion one has $r_3\!=\!1$. However, the 
$1/m_b$ corrections
are not negligible as is seen from the expression for the second
moment. On the other hand, it must be positive at any $m_b$. This
is the reason to expect that the power expansion for it is
sign-alternating, and that the $1/m_b^2$ correction
partially offsets the last term in Eq.~(\ref{50}). Therefore,
we use in practice $r_3=0.75\pm 0.25$ as an educated guess of these 
effects.

Following Refs.~\cite{misuse,uses} we estimate the biases in
Eqs.~(\ref{48}) using two ans\"atze for the heavy quark distribution function:
\bea
\nonumber
F_1(k_+)&=&N_1\,(\La\!-\!k_+)^\alpha \,e^{ck_+} \:\theta(\La\!-\!k_+),
\msp{5} \\
F_2(k_+)&=&N_2\,(\La\!-\!k_+)^\beta \,e^{-d(\La\!-\!k_+)^2}\:
\theta(\La\!-\!k_+) \; ,
\label{52}
\eea
with the parameters fixed by Eqs.~(\ref{50}). We will see that the two 
classes of functions, despite
their different functional dependence, yield very similar results 
once Eqs.~(\ref{50}) are imposed fixing their mean  and
variance. Therefore we take their average as the central value, and 
include the difference into the estimate of the theoretical uncertainty.

\begin{figure}[t]
\mbox{\psfig{file=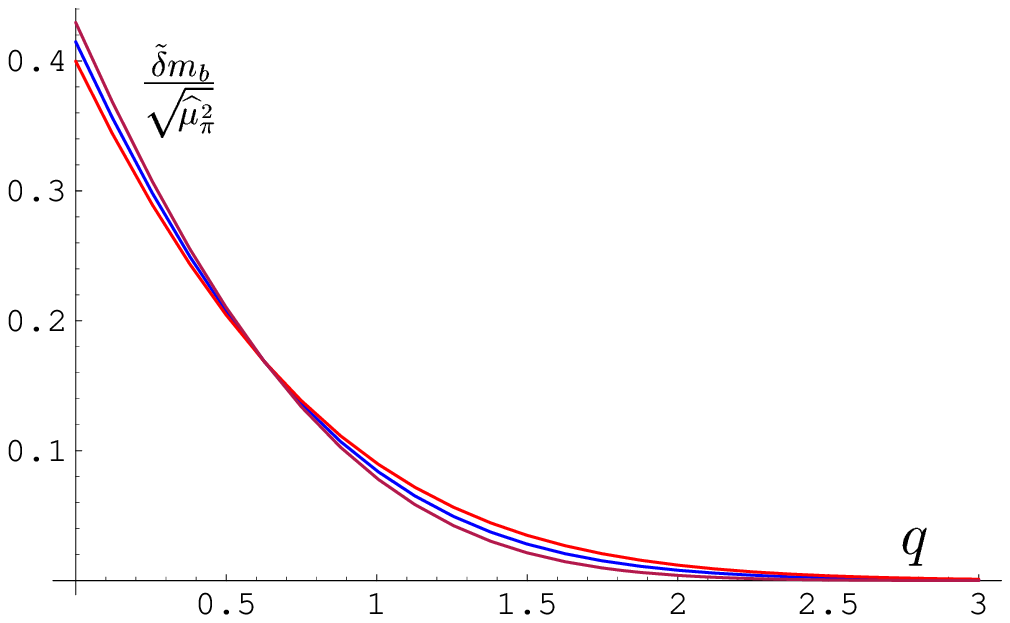,width=75mm}}
\hfill 
  \mbox{\psfig{file=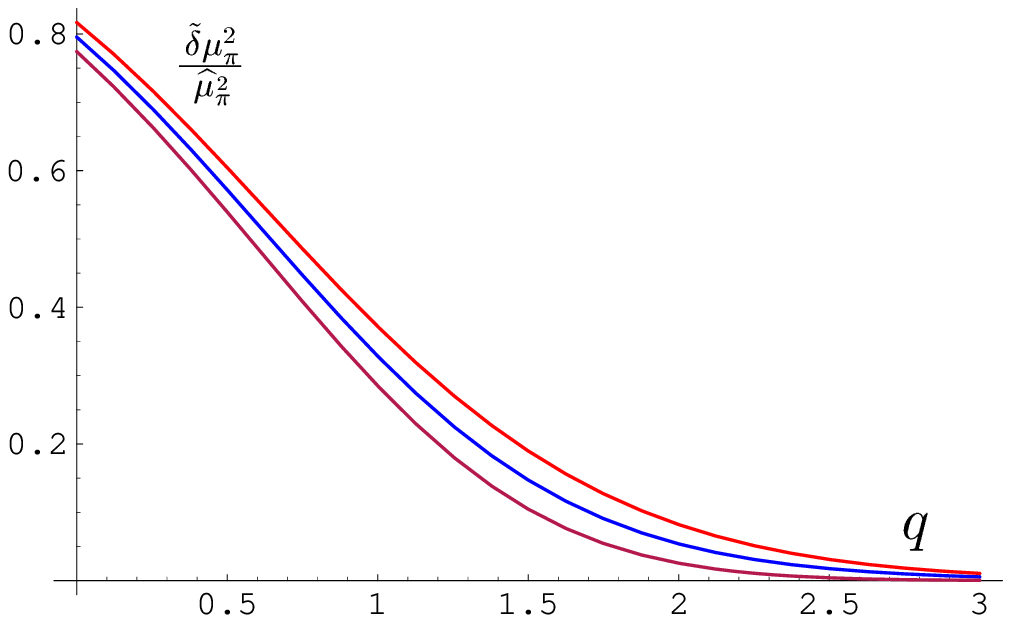,width=75mm}}
\caption{{\small
The dimensionless ratios
$\tilde \delta m_b/(\widehat{\mu_{\pi}^2})^{\frac{1}{2}}$
and $\tilde\delta \mu_{\pi}^2/\widehat{\mu_{\pi}^2}$ as functions of
the cut scale variable
$q\!=\!\hat Q/(\widehat{\mu_{\pi}^2})^{\frac{1}{2}}$
as described in the text. The middle blue curve gives the central value, 
the average of the biases obtained using the two distribution
functions. } }
 \label{MBMUPIBIAS}
 \end{figure}

In general, the adopted distribution functions depend on the values of
${m_b}$ and ${\mu_\pi^2}$,
and the biases additionally on $E_{\rm cut}$. Within the
$1/m_b$ expansion the latter depend primarily on the combination
$\hat Q\!=\!\widehat{m_b}\!-\!2E_{\rm cut}$ (directly related to the hardness
${\cal Q}\!\simeq\! M_B\!-\!2E_{\rm cut}$) rather than on $E_{\rm cut}$ and
$m_b$ separately. By dimensional arguments
they are actually functions of two variables since the absolute scale
$m_b$ does not enter. Yet even with these simplifications the resulting 
expressions are cumbersome, and it would not be enlightening to 
present them here. 
Fortunately it is not
necessary; for we have verified that the biases follow simple scaling 
behavior   
with reasonable accuracy sufficient for our purposes.
Once calibrated by
$\sqrt{\widehat{\mu_{\pi}^2}}$, the biases
depend primarily on
the ratio $q\equiv \hat Q/\sqrt{\widehat{\mu_{\pi}^2}}\,$:
\beq
\frac{\tilde\delta m_b}{\sqrt{\widehat{\mu_{\pi}^2}}}
=f\left(q,r\right)\simeq f_s\left(q\right)\; , \qquad
\frac{\tilde\delta \mu_{\pi}^2}{\widehat{\mu_{\pi}^2}}=g\left(q,r\right)
\simeq g_s\left(q\right)\;,
\label{54}
\eeq
with the residual dependence on the remaining ratio
$\,r\!=\!\widehat{\mu_{\pi}^2}/\widehat {\Lambda}^2$ rather weak 
($\widehat {\Lambda}\!\equiv\! M_B\!-\!\widehat{m_b}$). The
functions $f_s$ and $g_s$ are shown in
Figs.~\ref{MBMUPIBIAS}.
(For illustration in Sect.~5 we also give the corresponding 
estimate for the third moment of the distribution.) 
The difference between the two curves which represent the results for 
the two classes of functions
defined in Eq.~(\ref{52}) gives a sense of the scale of the theoretical 
uncertainty, albeit `cum grano salis':
having them coincide at some point does of course not mean there is 
then no uncertainty. The functions $f_s(q)$ and $g_s(q)$ can adequately 
be described by the following simple functions for realistic values
of $q$ and $r$: 
\beq
f_s(q)\simeq 0.415 \,\exp{[-1.3\,q-0.28\,q^{2.5}]}\,,\qquad 
g_s(q)\simeq 0.80 \,\exp{[-0.60\,q-0.27\,q^{2.5}]}\;.
\label{64}
\eeq
 
 \subsection{Biases and perturbative corrections}
 
The above bias evaluation was based purely on the nonperturbative
primordial distribution. In reality cuts are applied to the observable
spectrum resulting from the convolution of the perturbative and the
nonperturbative pieces. Perturbative corrections modify the
bias. Moreover, the effect of the perturbative corrections depends on
their implementation, in particular on the separation scale
$\mu$. Indeed, the moments of the combined spectrum are
$\mu$-independent at a given cut, as it has to be for an
observable. The biases computed with the nonperturbative distribution
alone, on the other hand, would depend on the choice of $\mu$ since
the distribution function and, in particular $m_b$ depend on the
normalization scale.

It has been observed \cite{misuse} that with the prototype Wilsonian
perturbative spectrum of Ref.~\cite{uses} perturbative contributions
had little impact on the biases. 
Having calculated the perturbative spectrum explicitly with 
the Wilsonian cutoff, in this paper we adopt a more accurate direct 
approach. Namely, we determine the complete spectrum starting from
this Wilsonian perturbative kernel. Simultaneously, we
confront the results with the
corresponding OPE predictions which include complete expressions 
for the perturbative moments derived from this spectrum, 
see Appendix \ref{WILSPECT}.

\begin{figure}
\mbox{\psfig{file=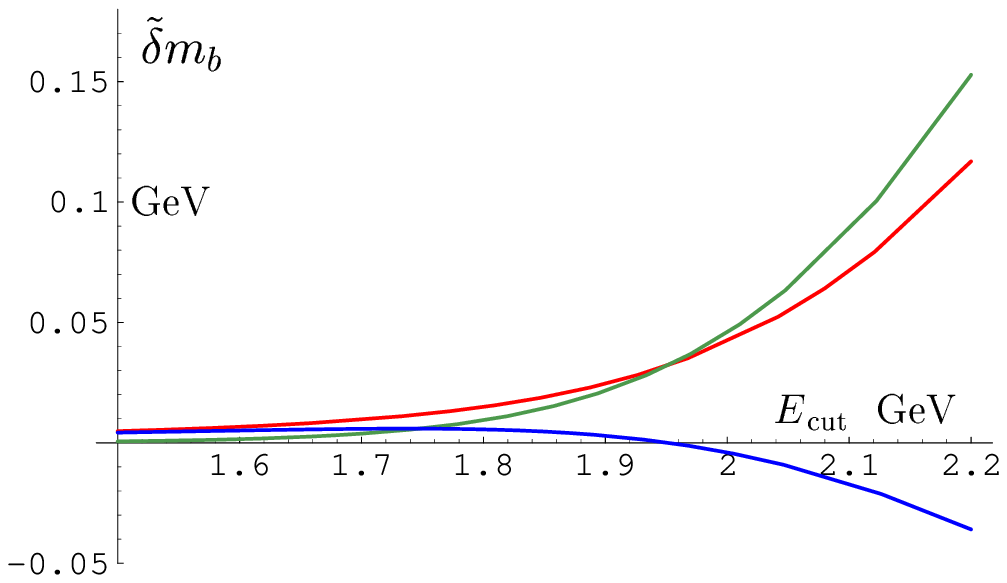,width=75mm}}
 \hfill 
 \mbox{\psfig{file=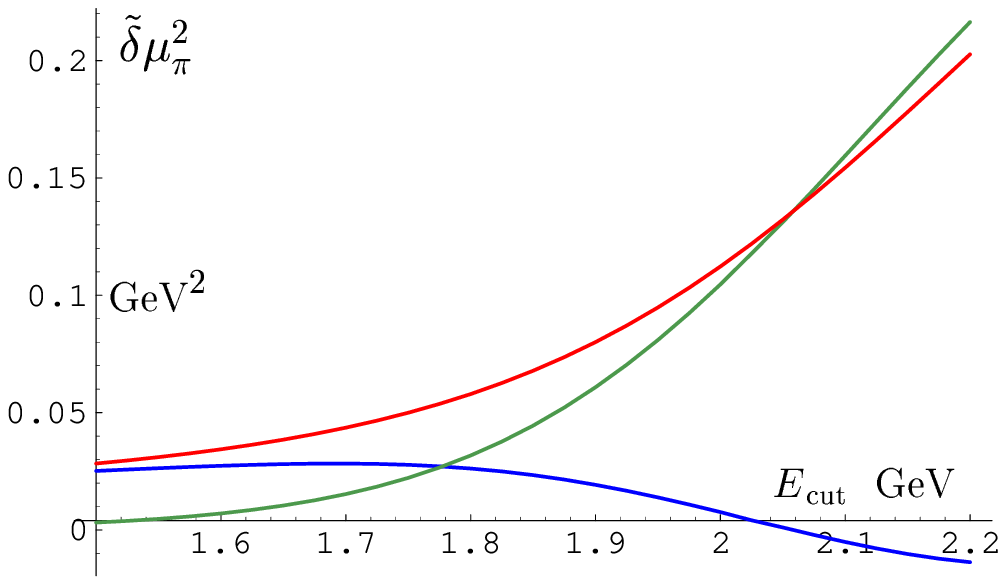,width=75mm}}
 \caption{{\small
The perturbative effects on the biases $\tilde\delta m_b$ and
$\tilde\delta\mu_{\pi}^2$. 
The red curve is the complete bias using the full spectrum. The
green curve is the bias calculated without perturbative corrections, 
and the blue line shows the difference between the two curves.}}
 \label{CONVBIAS}
 \end{figure}

Again we find that for $\mu \!\sim\! 1\GeV$ the biases are only weakly
modified by the perturbative corrections. 
This is illustrated by Figs.~\ref{CONVBIAS}, where we have 
assumed $m_b(1\GeV) \!=\! 4.6\GeV$ and 
$\mu_{\pi}^2(1\GeV)\!=\! 0.45\GeV^2$.  
Even so in our analysis we retain the perturbative shifts for the sake
of completeness.  At the same time, due to their smallish size
we just approximate these shifts in $\tilde\delta m_b$ and in $\tilde\delta
\mu_\pi^2$ by functions of only the single variable  $q$, obtained at the
central values expected for the parameters: 
\bea
\nonumber
\Delta^{\rm pert} f_s(q) \msp{-3} &\simeq & \msp{-2}
0.0103\,q^3\!-\!0.0729\,q^2\!+\!0.165\,q\!-\!0.111 \\
\Delta^{\rm pert} g_s(q) \msp{-3}&\simeq & \msp{-4}
-0.0138\,q^3\!+\!0.0.0437\,q^2\!+\!0.0316\,q\!-\!0.042 \qquad
\label{84}
\eea
If necessary, these can be
calculated anew at the final fitted values; the procedure converges
well since the shifts are small.  
 
\subsection{Procedure}

Let us summarize now the proposed way to obtain the numerical theory
predictions for the photon energy moments. The inputs  are 
the (true) values
of the heavy quark parameters; they are assumed to be normalized at 
$\mu\!=\!1\GeV$. The procedure consists of three steps. First, the `biased',
i.e.\ uncorrected moments are evaluated by adding the
nonperturbative and perturbative contributions. The former are 
given by Eqs.~(\ref{40}), (\ref{41}). The perturbative ones are
described in detail in Appendix~A.1; they refine
the previously applied Eqs.~(\ref{21}).\footnote{The numerical
difference appears insignificant, though.} 
The first-order integrals $a_k$ are given in Eqs.~(\ref{231}) and 
(\ref{232}) with
$B_1$ defined in Eq.~(\ref{222}). The similar BLM pieces of the
moments, $b_k$, are obtained by integrating Eq.~(\ref{250})
(and, for $b_1$, additionally subtracting the value of 
$B_2$ given by Eq.~(\ref{251})). Alternatively, one can use the 
approximate numerical interpolations Eqs.~(\ref{282}) and (\ref{283}).

Next with $\widehat{m_b}$ and
$\widehat{\mu_{\pi}^2}$ defined in Eq.(\ref{50}) we can determine the
biases $\tilde \delta m_b$ and $\tilde \delta \mu_{\pi}^2$ using
Eq.~(\ref{54}):
\beq
\tilde\delta m_b=\sqrt{\widehat {\mu_\pi^2}} \, f_s(q)\,, \qquad
\tilde\delta\mu_\pi^2 = \widehat {\mu_\pi^2} \, g_s(q)
\label{82}
\eeq
with the dimensionless scaling variable 
$q=\!\hat{Q}/\sqrt{\widehat{\mu_\pi^2}}\,$, where 
$\hat Q \!=\! \widehat{m_b} \!-\! 2E_{\rm cut}$.

Finally, we may apply additional small adjustments accounting for the
perturbative impact on the biases. While obtained in a lengthy way
using  complicated expressions, they
can be approximated quite adequately by simple polynomials as in
Eqs.~(\ref{84}).
To account for these effects one would replace, at the previous step 
\beq
f_s(q) \Longrightarrow f_s(q)+\Delta^{\rm pert} f_s(q), \qquad  
g_s(q) \Longrightarrow g_s(q)+\Delta^{\rm pert} g_s(q)\;.
\label{86}
\eeq
The thus corrected photon energy moments can then be confronted with the 
data.
\vspace*{1mm}

Here we illustrate the procedure using the central values of the 
heavy quark parameters \cite{buchpriv} very close to those reported
by BaBar \cite{babarprl}:
\bea
m_b(1\GeV)\msp{-4}&=&\msp{-4}4.61 \GeV, \qquad 
\mu_\pi^2(1\GeV)=0.41 \GeV^2  \\
\rho_D^3(1\GeV)\msp{-4}&=&\msp{-4} 0.17\GeV^3.
\label{92} 
\eea
Likewise we use the central number for  
the chromomagnetic value  
inferred from the $B^*-B$ mass difference, and we fix the one 
for the $LS$ term to its theoretical expectation:
\beq 
\mu_G^2(1\GeV) = 0.35\GeV^2 , 
\qquad \rho_{LS}^3 =-0.15\GeV^3\;. 
\label{92a}
\eeq

For $E_{\rm cut}\!=\!2.0 \GeV$ we have  $x\!=\!0.8677$ and for the
perturbative contributions we find from Eqs.~(\ref{231}), (\ref{232}),
(\ref{260})--(\ref{251}) and (\ref{302}):
\bea
\nonumber
a_1(0.8677;\mbox{$\frac{1}{4.61}$})+\hat a_1(0.8677)=-0.0511\,,
& & b_1(0.8677;\mbox{$\frac{1}{4.61}$})+\hat b_1(0.8677)=-0.0909\\
a_2(0.8677;\mbox{$\frac{1}{4.61}$})+\hat a_2(0.8677)=\msp{-.1} 0.00447\,,
& & b_2(0.8677;\mbox{$\frac{1}{4.61}$})+\hat b_2(0.8677)=0.00538\,; \qquad
\label{492}
\eea
using $\alpha_s\!=\!0.22$ and $\beta_0\!=\!9$ yields
\beq
\aver{ E_{\gamma}}^{\rm pert}|_{E_{\gamma}>2.0 \,{\rm GeV}}=2.328 \GeV, \qquad
\aver{E_\gamma^2\!-\!\aver{E_\gamma}^2}^{\rm pert}|
_{E_{\gamma}>2.0 \,{\rm GeV}}=0.00390 \GeV^2.
\label{494}
\eeq

The nonperturbative power corrections given by
Eqs.~(\ref{40}) and (\ref{41}) are  
\beq
\aver{E_{\gamma}}^{\rm power}=-0.00715 \GeV, \qquad 
\aver{E_{\gamma}^2\!-\!\aver{E_{\gamma}}^2}^
{\rm power}=0.0253 \GeV^2.
\label{495}
\eeq
Adding the perturbative corrections and these power corrections
results in
\beq
\aver{ E_{\gamma}}|_{2.0\,{\rm GeV}}^{\rm biased}= \,2.321 \GeV \,, \qquad
\aver{(E_{\gamma}\!-\!\aver{E_{\gamma}})^2}|_{2.0\,{\rm GeV}}^{\rm biased} 
= \,0.0292 \GeV^2\;.
\label{496}
\eeq

The last  step is to correct for the biases. For our set of
parameters, Eqs.~(\ref{50}) yield $\widehat{m_b}\!=\!4.602 \GeV$ or
$\widehat{\Lambda}\!=\!0.677\GeV$ and, using $r_3\!=\!0.75$, we obtain 
$\widehat{\mu_{\pi}^2} \!=\! 0.330 \GeV^2$.  
At $E_{\rm cut}\!=\!2.0 \GeV$ the value of the effective normalized
hardness is then $q\!=\!\hat {Q}/\sqrt{\widehat{\mu_\pi^2}}
\!\simeq\! 1.05 $ and we
have $r\!=\!\widehat{\mu_\pi^2}/\widehat\Lambda^2\!\simeq\!0.72$.
We take the scaling form Eqs.~(\ref{82}) for the biases parametrized
in Eqs.~(\ref{64}), and apply the shifts Eqs.~(\ref{84}) due to 
the perturbative effects according to  Eqs.~(\ref{86}):
\bea
\nonumber
\tilde\delta m_b \msp{-3} &=& \msp{-3} \sqrt{\widehat {\mu_\pi^2}} \left(\,
f_s(q)+\Delta^{\rm pert} f_s(q)\right)=0.041 \GeV\\ 
\tilde\delta\mu_\pi^2 \msp{-3}&=&\msp{-3} \widehat {\mu_\pi^2} \left(\,
g_s(q)+\Delta^{\rm pert} g_s(q)\right)=0.111 \GeV^2 \;.
\label{497}
\eea

We arrive at the final bias-corrected values of the moments combining
Eqs.~(\ref{496}) and (\ref{497}) according to Eqs.~(\ref{48}): 
\bea
\nonumber
\aver{ E_{\gamma}}|_{2.0\,{\rm GeV}} \msp{-4}&=& \msp{-4}
\aver{ E_{\gamma}}|_{2.0\,{\rm GeV}}^{\rm
biased}+\mbox{$\frac{\tilde\delta m_b}{2}$}=2.342  \GeV\\ 
\aver{(E_{\gamma}\!-\!\aver{E_{\gamma}})^2}|_{2.0\,{\rm GeV}} 
\msp{-4}&=&\msp{-4}
\aver{(E_{\gamma}\!-\!\aver{E_{\gamma}})^2}|_{2.0\,{\rm GeV}}^{\rm biased} -
\mbox{$\frac{\tilde\delta \mu_\pi^2}{12}$}=0.020\GeV^2 \;.
\label{499}
\eea 
\vspace*{1.5mm}

If we follow the similar procedure for other cuts, we find, 
for the biased average photon energy and the variance,
the following values for 
$E_{\rm cut}\!=\!1.8$, $1.9$, $2.0$ and $2.1 \GeV$:
\bea
\nonumber 
\aver{ E_{\gamma}}|_{1.8\,{\rm GeV}}^{\rm biased} 
\msp{-4}&\simeq&\msp{-4} 2.305  \GeV 
\qquad \qquad  
\aver{(E_{\gamma}\!-\!\aver{E_{\gamma}})^2}|_{1.8\,{\rm GeV}}^{\rm biased}
\simeq 0.0357 \GeV^2 \\
\nonumber 
\aver{ E_{\gamma}}|_{1.9\,{\rm GeV}}^{\rm biased} 
\msp{-4}&\simeq&\msp{-4} 2.313  \GeV 
\qquad \qquad  
\aver{(E_{\gamma}\!-\!\aver{E_{\gamma}})^2}|_{1.9\,{\rm GeV}}^{\rm biased}
\simeq 0.0321\GeV^2 \\ 
\nonumber 
\aver{ E_{\gamma}}|_{2.0\,{\rm GeV}}^{\rm biased} 
\msp{-4}&\simeq&\msp{-4} 2.321 \GeV 
\qquad \qquad  
\aver{(E_{\gamma}\!-\!\aver{E_{\gamma}})^2}|_{2.0\,{\rm GeV}}^{\rm biased}
\simeq 0.0293 \GeV^2 \\
\aver{ E_{\gamma}}|_{2.1\,{\rm GeV}}^{\rm biased} 
\msp{-4}&\simeq&\msp{-4} 2.329   \GeV 
\qquad \qquad  
\aver{(E_{\gamma}\!-\!\aver{E_{\gamma}})^2}|_{2.0\,{\rm GeV}}^{\rm biased}
\simeq 0.0271  \GeV^2 \;. 
\label{96}
\eea 
The bias corrections read 
\bea
\nonumber 
\tilde\delta m_b|_{1.8\,{\rm GeV}} \msp{-4}&\simeq &\msp{-4} 
0.015  \GeV\, \qquad \qquad  
\tilde \delta \mu_{\pi}^2|_{1.8\,{\rm GeV}} \simeq 0.0574 \GeV^2\\ 
\nonumber 
\tilde\delta m_b|_{1.9\,{\rm GeV}} \msp{-4}&\simeq &\msp{-4}  
0.024 \GeV\, \qquad \qquad 
\tilde \delta \mu_{\pi}^2|_{1.9\,{\rm GeV}} \simeq  0.0790 \GeV^2 \\
\nonumber 
\tilde\delta m_b|_{2.0\,{\rm GeV}} 
\msp{-4}&\simeq &\msp{-4} 0.041  \GeV\, \qquad \qquad 
\tilde \delta \mu_{\pi}^2|_{2.0\,{\rm GeV}} \simeq  0.111 \GeV^2 \\
\tilde\delta m_b|_{2.1\,{\rm GeV}}\msp{-4} 
&\simeq & \msp{-4} 0.070  \GeV\, \qquad \qquad 
\tilde \delta \mu_{\pi}^2|_{2.1\,{\rm GeV}} \simeq  0.152 \GeV^2 . 
\qquad \label{98}
\eea
Then we arrive at
the following final, i.e.\ bias-corrected predictions for the 
average and variance,
\bea 
\nonumber
\aver{ E_{\gamma}}|_{1.8\,{\rm GeV}} \msp{-4}&\simeq &\msp{-4} 2.312 \GeV,  
\qquad \qquad 
\aver{(E_{\gamma}\!-\!\aver{E_{\gamma}})^2}|_{1.8\,{\rm GeV}} \simeq 0.0309 
\GeV^2 \\
\nonumber 
\aver{ E_{\gamma}}|_{1.9\,{\rm GeV}} \msp{-4}&\simeq &\msp{-4} 2.325  \GeV,  
\qquad \qquad 
\aver{(E_{\gamma}\!-\!\aver{E_{\gamma}})^2}|_{1.9\,{\rm GeV}} \simeq 0.0255
\GeV^2\\
\nonumber 
\aver{ E_{\gamma}}|_{2.0\,{\rm GeV}} \msp{-4}&\simeq &\msp{-4} 2.342  \GeV,  
\qquad \qquad 
\aver{(E_{\gamma}\!-\!\aver{E_{\gamma}})^2}|_{2.0\,{\rm GeV}} \simeq 0.020
\GeV^2\\  
\aver{ E_{\gamma}}|_{2.1\,{\rm GeV}} \msp{-4}&\simeq &\msp{-4} 2.364  \GeV ,  
\qquad \qquad 
\aver{(E_{\gamma}\!-\!\aver{E_{\gamma}})^2}|_{2.1\,{\rm GeV}} \simeq  0.0145
\GeV^2 .
\label{102} 
\eea

These numbers, in principle, can be compared with some published 
measurements (a more detailed analysis  of the BELLE data has also been
performed\,\footnote{O.~Buchmueller, 
private communication}\,):  
\bea 
\nonumber
\aver{ E_{\gamma}}|_{1.8\,{\rm GeV}} \msp{-4}&\simeq&\msp{-4} 2.292 \pm 0.026 
\pm 0.034 \; \GeV  
\; \; \; \msp{9.7} \mbox{BELLE } \cite{BELLE} \\
\nonumber
\aver{(E_{\gamma}\!-\!\aver{E_{\gamma}})^2}|_{1.8\,{\rm GeV}} 
\msp{-4}&\simeq&\msp{-4} 0.0305 \pm 
0.0074 \pm 0.0063 \; (\GeV )^2 
\; \; \; \mbox{BELLE } \cite{BELLE}  \\
\nonumber
\aver{ E_{\gamma}}|_{2.0\,{\rm GeV}} \msp{-4}&\simeq&\msp{-4} 
2.346 \pm 0.032 \pm 0.011 \; \GeV  
\; \; \; \msp{11.9} \mbox{CLEO } \cite{CLEO} \\ 
\aver{(E_{\gamma}\!-\!\aver{E_{\gamma}})^2}|_{2.0\,{\rm GeV}} 
\msp{-4}&\simeq&\msp{-4} 0.0226 \pm 
0.0066 \pm 0.0020 \; (\GeV )^2 
\; \; \; \msp{.9} \mbox{CLEO } \cite{CLEO} 
\label{103}
\eea
In view of the uncertainties in the experimental numbers and their
high correlations we refrain from detailed comparison; yet there
appears a clear trend that the bias corrections bring the predictions
into close agreement with the data, in particular for the variance.
\vspace*{1.5mm}

We want to emphasize the following features in the  estimated values for
biases:\vspace*{-1.5mm} 
\begin{itemize}
\item 
The biases increase steeply with the energy cuts.\vspace*{-1.5mm} 
\item 
While for $E_{\rm cut}\!=\! 1.8 \GeV$ they are rather insignificant, 
for  $E_{\rm cut} \!\geq\! 2.0 \GeV$ they become comparable 
to or even larger than the accuracy with which the corresponding 
parameters have been be extracted from $B \tto X_c\,\ell\nu$.\vspace*{-1.5mm} 
\item 
The uncertainties inherent in the evaluation of biases remain rather
small for $m_b$, and even for $\mu_{\pi}^2$ they are of moderate size. 
\end{itemize}

\section{Effect of the four-quark operators at order
\boldmath $\alpha_s/m_b^3$}
\label{4QUARK}


As noted previously \cite{motion,optical,uses},
four-quark operators
$(\bar{b}\Gamma _i s)\, (\bar{s}\Gamma _i b)$ do not contribute to 
$b\tto s\!+\!\gamma$ at tree level, unlike for
$b\tto u \,\ell \nu$. Yet hard gluon emission induces Wilson coefficients 
for them already at order $\alpha_s$. Here we give their contributions.
Similar short-distance corrections affect, in general, 
Wilson coefficients of all other
operators, including those determining the moments of the light-cone
distribution function. The four-quark effects, however are usually 
numerically enhanced, at least where they are allowed at the tree level. 

To order $\alpha_s$ the effect is expressed in terms of two
expectation values
\bea
\nonumber
S_V \msp{-4}&=&\msp{-4}
(\delta_{ik}\delta_{lj}\!-\!\mbox{$\frac{1}{3}$}\delta_{ij}\delta_{lk})
\frac{1}{2M_B}\,
\matel{B}{\bar{b}^i\gamma_\mu b^k\, \bar{s}^l\gamma^\mu s^j }{B}\\
S_A\msp{-4}&=&\msp{-4}
(\delta_{ik}\delta_{lj}\!-\!\mbox{$\frac{1}{3}$}\delta_{ij}\delta_{lk})
\frac{1}{2M_B}\,
\matel{B}{\bar{b}^i\gamma_\mu\gamma_5 b^k
\bar{s}^l\gamma^\mu \gamma_5 s^j }{B} \; ,
\label{310}
\eea
where the $i,j,k,l$ denote color indices, and yields the following 
contributions to the first three moments:
\bea
\nonumber
\Delta_{\bar{s}s}\aver{(E_\gamma\!-\!\aver{E_\gamma})^2}
\msp{-4}&=&\msp{-4} \!\!-\frac{\pi\alpha_s}{4} 
\left(\frac{S_V}{m_b}-\frac{1}{3}\frac{S_A}{m_b}\right)\\
\nonumber
\Delta_{\bar{s}s}\aver{E_\gamma}
\msp{-4}&=&\msp{-4} \;\;\;\frac{\pi\alpha_s}{2} \left(-\frac{S_V}{m_b^2}+
3\frac{S_A}{m_b^2}\right)\\
\frac{1}{\Gamma_{B\to X\gamma}}\Delta_{\bar{s}s}\Gamma_{B\to X\gamma} 
\msp{-4}&=&\msp{-4} \;\;\;2\pi\alpha_s \left(\frac{S_V}{m_b^3}+
\frac{11}{3}\frac{S_A}{m_b^3}\right)\;;
\label{312}
\eea
the appropriate value of $\alpha_s$ here is about
$0.3$. $S_{V,A}$ are related to the non-valence four-quark
parameters $\omega_{V,A}$, $\tau_{V,A}$ of Ref.~\cite{four}
through 
\beq
S_V = \mbox{$\frac{8}{9}$}\omega_V -\mbox{$\frac{2}{3}$}\tau_V\,,
\qquad
S_A =
\mbox{$\frac{8}{9}$}\omega_A -
\mbox{$\frac{2}{3}$}\tau_A\;.
\label{314}
\eeq

It should be noted that the contributions from the four-quark operators 
are $1/m_b^3$-suppressed
only in the total width. In the higher moments an enhancement arises 
over naive expectations one might infer considering the fact that the
four-quark operators contribute to the spectrum only in the 
end-point region where $E_\gamma\!-\!\frac{m_b}{2}\!\sim\!\mhad$, 
and that their overall effect on the width is  $1/m_b^3$. The 
diagrams with a gluon
exchanged between the light quark legs generate additional
powers of $\epsilon\!=\!m_b\!-\!2E_\gamma$ in the denominator in the
forward scattering amplitude; this directly affects the higher
moments, but not the total rate. Accordingly, we find the
corrections to the rate negligible, yet not necessarily so for
$\aver{E_\gamma}$ and, in particular, for the second moment
determining $\mu_\pi^2$. Since contributions from $(\bar b \Gamma_i 
s)(\bar s \Gamma _i b)$ can
arise only via non-valence configurations in the $B$ meson (for the KM 
favored modes), they must be somewhat suppressed; yet the degree of 
suppression is uncertain.

The impact on the second moment and, therefore
the related shift in the apparent value of $\mu_\pi^2$ 
could a priori be significant -- if the generic non-valence
`color-straight' operators (in the terminology of Ref.~\cite{four}) are 
estimated through
the vacuum chiral condensate $\aver{\bar{s}s}_0$ according to
\beq
\frac{1}{2M_B}\,
\matel{B}{\bar{b}^i b^i\, \bar{s}^k s^k(0) }{B} \sim
\frac{1}{2M_B}\,
\matel{B}{\bar{b}^i b^i(0)}{B}\,\aver{\bar{s}s}_0 \simeq
-(200\MeV)^3\;.
\label{320}
\eeq
A dedicated discussion of the non-factorizable four-fermion
expectation values can be found in Ref.~\cite{four}. It
turns out, though that the value of $S_V$ is expected
to be particularly suppressed compared to the estimate
(\ref{320}). $S_V$, to leading order in $1/N_c$,
reduces to the color-straight vector operator,
non-valence expectation values for which (giving the
$s$-quark number density at  the origin) are probably small, since
the density must vanish identically when integrated over the
whole volume. The $1/N_c$ color-octet piece of $S_V$
contributing to the Darwin expectation value $\rho_D^3$, is
probably larger, yet still suppressed.

A larger value is expected for the spin-dependent axial
matrix element $S_A$, for which Eq.~(\ref{320}) can be
conservatively taken as an upper bound. Yet the coefficient for 
$S_A$ is appreciable
only in $\aver{E_\gamma}$ and in the total width where the effects
are $1/m_b^2$- and $1/m_b^3$-suppressed,
respectively. Therefore we think that the related
corrections in $\aver{E_\gamma}$ are below $10\MeV$ in $m_b$ and 
safely at the sub-percent level for the total width.
\vspace*{2mm}

\section{Cut-related uncertainties in \boldmath $\;b\!\to\!c\,\ell\nu$}
\label{SLBIAS}

Lower cuts imposed on the charged lepton energy in $B\tto X_c
\,\ell\nu$ decays likewise degrade the hardness of the transition. Yet
the final state in $b\to c \ell \nu$ exhibits kinematics more complex
than in $b\to s + \gamma$, which makes this effect less
transparent. Furthermore the dynamics is considerably more complicated
as well, since the relevant distribution describing the dependence on
the charged lepton energy cut, generally is neither the light-cone
distribution nor the temporal distribution describing the effects of
the `Fermi motion' in the small velocity (SV) kinematics. There is no
reason to expect an analogous distribution to be universal for
different moments; the power corrections are probably significant in
the realistic settings, and probably vary depending on the moments as
well.

Nevertheless there are some common features between $B\tto X_s+\gamma$
and $B\tto X_c \,\ell\nu$, in particular due to the conceptual
similarity of the OPE treatment of both cases. Furthermore the full
impact of the degraded hardness might not reveal itself in the few
lowest terms in the $1/m_b$ and $\alpha_s$ expansion  that one is, in
practice, limited to in the theoretical description. 
 
As argued in Ref.~\cite{misuse}, in this situation it is reasonable to
adopt a cautious approach and use the bias estimates for $B\tto
X_s+\gamma$ merely to assess the potential magnitude of {\sf
additional} cut-related uncertainties in the theoretical predictions
for the $b\tto c \,\ell\nu$ transitions. Namely, the conventional OPE
predictions for a particular moment, with or without the cut, depend
on the heavy quark parameters. Assigning a cut-induced 'uncertainty'
to the values of those parameters (even though they have definite
values in actual QCD), would result in an additional variation in the
theoretical predictions, which can be used as an estimate of the scale
of possible cut-related exponential pieces.
 
To implement this prescription in practice, we need a commensurate
definition for the hardness in both transitions and how that  scale is
degraded by energy cuts. We advocate the following ansatz which is more
accurate at low hardness:
\beq
{\cal Q}_{\gamma} \simeq M_B\!-\!E_{\rm cut}^\gamma\!-
\!\sqrt{(E^\gamma_{\rm cut})^2+M_{K^*}^2}
\qquad \mbox{\sf vs.} \qquad
{\cal Q}_{\rm sl} \simeq M_B\!-\!E_{\rm cut}^\ell\!-\!\sqrt{(E^\ell_{\rm 
cut})^2+M_{D^*}^2} \; ;
\label{133}
\eeq
equating the two yields the prescription
\beq
E_{\rm cut}^{\gamma} \Longleftrightarrow E_{\rm cut}^{\rm sl} +
\frac{M_{D^*}^2-M_{K^*}^2}
{2(E_{\rm cut}^{\rm sl}\!+\!\sqrt{(E^{\rm sl}_{\rm cut})^2\!+\!M_{D^*}^2})}
\;.
\label{134}
\eeq
for identifying equivalent energies in radiative and semileptonic
decays. The resulting dependence is practically linear,
Fig.~3. 
\begin{figure}[h]
\begin{center}
\mbox{\psfig{file=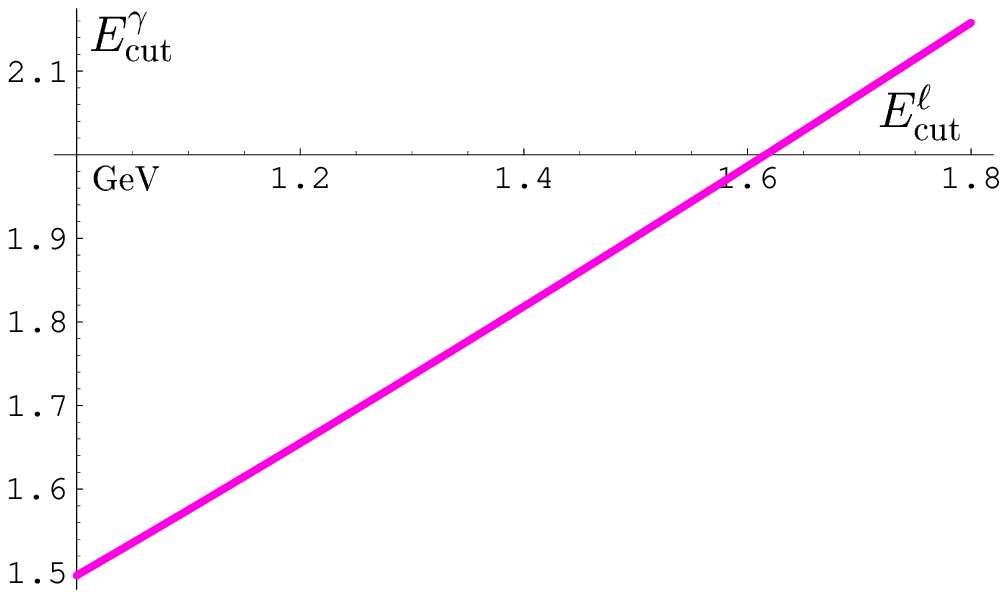,width=70mm}}
\caption{{\small
Cuts 
in $B\tto X_s\!+\!\gamma\,$ and in $B\tto X_c\,\ell\nu\,$ yielding
the same overall hardness ${\cal Q}$
} }\vspace*{-20pt}
\end{center}
\end{figure}
For a given $E_{\rm cut}^{\rm sl}$ we use Eq.~(\ref{134}) to
determine the corresponding $E_{\rm cut}^{\gamma}$. This allows us 
 to estimate 
the size of biases associated with $m_b$, $\mu_\pi^2$, $\rho_D^3$,
... at this hardness. Using the explicit dependence of the
semileptonic moments in question on these heavy quark parameters, one
can estimate the degree to which the theoretical accuracy deteriorates
due to the insufficient energy release. 
It is interesting to note that
Refs.~\cite{ckm03fpcp} concluded that when the lepton cut is raised to
$1.7\GeV$ theory looses control over $\aver{M_X^2}$. According
to Fig.~3 the lepton cut at $1.7\GeV$ corresponds to the cut on
$E_\gamma$ at about $2.1\GeV$ -- and this is just the limit where the OPE 
for $B\tto X_s\!+\!\gamma$ totally breaks down, according to the 
present estimates.

A possible physical motivation behind this procedure has been
mentioned in Ref.~\cite{misuse}: the lower cuts both on $E_\gamma$ or
on $E_\ell$ kinematically reduce or even eliminate the contribution of
high mass hadronic final states in the decay, thus introducing an
effective `normalization scale' $\mu\!\approx \!{\cal Q}$. By virtue
of the heavy quark sum rules, the integrals of the inclusive decay
probabilities yield the values of the nonperturbative heavy quark
parameters, with the upper cutoff on the mass of the produced states
playing the role of the normalization point for the operators
\cite{optical}. For large enough values of the hardness, 
explicit perturbative
corrections account for this scale dependence, yet the possible
nonperturbative dependence at limited ${\cal Q}$ is missed.  This
ansatz can therefore be interpreted as accounting for a certain
nonperturbative evolution of the heavy quark parameters at relatively
low normalization point $\mu$, provided it is to some extent
universal, similar to the universality of the perturbative
$\mu$-dependence of a given nonperturbative expectation value at $\mu$
in the perturbative domain.
\begin{figure}[h]
\vspace*{-3mm}
\begin{center}
\mbox{\psfig{file=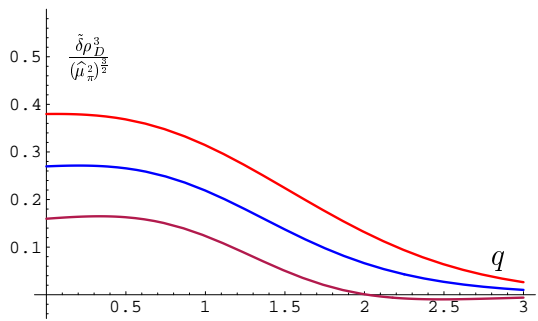,width=70mm}}
\vspace*{-3mm}
\caption{{\small
The expected bias for $\rho_D^3$: the dimensionless ratio $\tilde\delta 
\rho_D^3/(\widehat{\mu_{\pi}^2})^{\frac{3}{2}}$ for the two
distribution functions and their average (blue) \vspace*{-15pt} 
} }
\end{center}
\end{figure}

For the sake of completeness we illustrate in Fig.~4 the  expected
size of the similar bias $\tilde\rho_D^3$ in the Darwin expectation 
value, which
following Eqs.~(\ref{48}) is defined (up to a factor $-24$) 
as the cut-related deficit in the third moment of the photon
spectrum. As anticipated,  the predictions for the bias 
itself are less certain, yet we see that the corresponding effect can 
be significant down to rather low cuts.

\section{Conclusions}
\label{SUMMARY}

In the present paper we addressed in detail the theoretical evaluation
of the photon energy moments in $B\tto X_s\!+\!\gamma$ decays with lower
cuts on the photon energy, in terms of the underlying heavy quark 
parameters. These moments are particularly important in connection to
the precision determination of the latter. Two general aspects have to
be addressed in  calculating the moments -- perturbative contributions
which act as corrections, and the nonperturbative effects dominating
the measured low moments. We have improved the former by computing the
perturbative spectrum with the explicit Wilsonian infrared cutoff on
the low-energy gluon modes. This has been accomplished analytically
including the BLM corrections. 

We believe that this approach ensures the most accurate and
trustworthy results for the heavy flavor decays with the heavy quark
mass in a few GeV range, as it is for the actual beauty
hadrons. Extensions of the method including resummation of the Sudakov
radiation effects would be required for much heavier quarks, with 
masses exceeding $20\GeV$. 

Adding to what has been stated in Sect.~2.1, an essential element of
the uncertainty in
the numerical outcome of Refs.~\cite{neubbsg} is related to the
unphysical, 
somewhat loose~\footnote{The model-independent exact
results for its asymptotic behavior emphasized in 
Refs.~\cite{neubbsg} in fact stem from the chosen scheme 
which, in some respects, is unsatisfactory itself. Thus their physical
significance is doubtful.} shape function $S(\omega)$ arrived at in
SCET in place of the truly Wilsonian $F(k_+)$.
Refs.~\cite{neubbsg} suggested that the properties of $S(\omega)$
required for constraining its possible form, may be related to the true
Wilson heavy quark expectation values measured, say, in $B\tto
X_c\,\ell\nu$ decays. Yet in practice such relations suffer from poorly
controlled perturbative corrections and seem too crude. It is unclear
if their accuracy can realistically be improved. The advantage of
employing the heavy quark distribution function  $F(k_+)$ is that its
moments are directly related to the extracted
expectation values; thus there is no significant uncertainty here.

Accordingly, we view our calculations of the spectrum and its moments as
in all respects superior theoretically to the SCET-based evaluations
of Refs.~\cite{neubbsg} and as more accurate numerically. 
This conclusion is illustrated, for instance, by
Fig.~5 showing the perturbative (Wilsonian) photon spectrum (the
$O_7$ part) in 
$b \tto s \!+\!\gamma$.
\begin{figure}[h]
\vspace*{-3mm}
\begin{center}
\mbox{\psfig{file=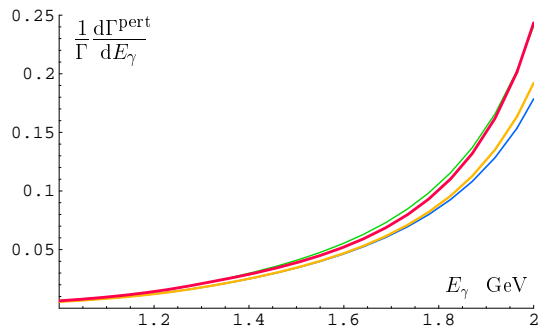,width=70mm}}
\caption{{\small
The significance of the higher-order BLM corrections
in the Wilsonian perturbative spectrum ($\mu\!=\!1\GeV$). Blue curve is the
first-order spectrum evaluated with $\alpha_s\!=\!0.35$. Yellow curve is
the usual $\alpha_s$ plus $\beta_0\alpha_s^2$ approximation, and the
(barely visible) green line is obtained by adding the
$\beta_0^2\alpha_s^3$ terms. Red curve shows the result of the
full BLM resummation which can be performed precisely in the Wilsonian
approach.\vspace*{-3mm}
}}
\end{center}
\end{figure}
It suggests that the fraction
of events with $E_{\gamma}\!<\! 1.8\GeV$ is not only small, but 
also reliably predicted, at least in what concerns the perturbative 
effects.

Doubts have been aired recently that the heavy-to-light transitions
may not be treated with the conventional OPE, even abstracting from
the complications associated with the high experimental cuts. In
particular, the principal OPE results established more than a decade
ago, were challenged. These results relate  the moments of, 
say the photon energy
spectrum to the local heavy quark expectation values which describe the
nonperturbative bound-state dynamics. The OPE ensures universality --
i.e.\ process-independence -- of these expectation values as it should be
for the characteristics of the initial bound state. Consequently, the
same expectation values can be measured, say through the moments of the
$B\tto X_c\,\ell\nu$ distributions.

We find such agnostic claims unfounded. The recent paper \cite{jet}
showed that, in spite of the well known peculiarities of the
light-like jet processes associated with the collinear bremsstrahlung,
the perturbative corrections to the moments of the photon spectrum are
calculable in the usual short-distance expansion, without encountering
new types of the nonperturbative effects. Our new perturbative
calculations implement this idea in practice. We state that the photon
moments {\sf can} be precisely calculated in terms of the analogous
expectation values determined from $B\tto X_c\,\ell\nu$ moments, as
long as the lower experimental cut on $E_\gamma$ is placed appropriately. 

On the nonperturbative side the so-called bias effects due to
significant lower cuts are in the focus of our study. They are missed
in the routinely used simplified OPE expressions, yet turn out
to be  significant in practice. 
We showed that for actual $B$ decays these biases are affected very
little by the perturbative corrections in the adequate Wilsonian
approach with the separation scale 
chosen  around the usual $1\GeV$. This
significantly simplifies incorporating the biases in practice. 
\vspace*{1.5mm}

The main phenomenological message can be expressed
rather concisely:\vspace*{-1.5mm}
\begin{itemize}
\item
Imposing a lower cut on the photon energy in $B\tto X_s + \gamma$ reduces 
the hardness ${\cal Q}$ of the transition.\vspace*{-1.5mm}
\item
This impacts the evaluation of the energy moments in ways that are not
fully described by the usual OPE expressions, thereby inducing  
systematic shifts, or biases.\vspace*{-1.5mm} 
\item
With the presently available accuracy of the measurements those biases 
can no longer be ignored.\vspace*{-1.5mm}
\item
While those biases cannot be evaluated in a completely
model-independent way yet, different ans\"atze for the heavy quark
distribution functions, when constrained by the OPE, yield a  
consistent picture:\vspace*{-1.5mm}
\begin{itemize}
\item
For $E_{\rm cut}^{\gamma} \!\lsim\! 1.7 \GeV$ they are more or less 
irrelevant. 
\item
For $E_{\rm cut}^{\gamma} \!> 2.1\!\GeV$ they are in all likelihood large, 
yet beyond accurate theoretical control.
\item
In the range $1.8\GeV \!\leq\! E_{\rm cut}^{\gamma} \!\leq\! 2$ GeV they 
are significant, yet can be estimated with reasonable accuracy. Thus 
they can be corrected for. \vspace*{-1.5mm}
\end{itemize}
\item
Lower cuts on the lepton energy in $B\tto X_c \,\ell \nu$ conceptually
have a similar impact on truncated lepton energy and hadronic mass
moments. Yet the resulting biases in the values of the moments cannot be
evaluated at present. We have suggested a simple prescription
to obtain semi-quantitative estimates of the potential uncertainties
associated with the high values of $E_{\rm cut}^\ell$.
\end{itemize}

We have presented the generally rather cumbersome  theoretical
evaluation of the truncated photon moments in a reasonably compact
form suitable even for global fits of various data on inclusive $B$
decays. In our opinion, such data analyses must be done by
experimentalists who can take care of numerous experimental
subtleties. As the numerical application of 
our analysis, we only note that we find good agreement between the
recently published BELLE data on the photon spectrum with the cut
as low as $1.8\GeV$, and the predictions based on the values of
$m_b$, $\mu_\pi^2$ and $\rho_D^3$ extracted by BaBar from $B\tto
X_c\,\ell\nu$ \cite{babarprl}. The data and theory predictions
coincide within $1\sigma$ of only the statistical error, cf.\ 
Eqs.~(\ref{102}) and (\ref{103}), even without invoking other error
bars or the uncertainties inherent in the theory calculations. 
\vspace*{2mm}

\noindent
{\bf Acknowledgments:} We are grateful to Th.~Mannel for discussions
and to our experimental colleagues U.~Langenegger, V.~Luth  and in 
particular to O.~Buchmueller for invaluable exchanges. Additional
comments on the difference in our treatment with the SCET emerged
after informative discussions with M.~Neubert at the Belle Workshop
and at KEK and his talks there, which are gratefully acknowledged.
This work was supported in part by the NSF under grant number PHY-0087419.

\renewcommand{\theequation}{A.\arabic{equation}}
\renewcommand{\thesection}{}
\setcounter{equation}{0}
\setcounter{section}{0}
\setcounter{table}{0}
\section{Appendices}
\renewcommand{\thesubsection}{A.\arabic{subsection}}

\subsection{Perturbative spectrum with the Wilsonian scale separation}
\label{WILSPECT}

In Ref.~\cite{imprec} it has been described in detail how the required
Wilsonian separation of soft and hard gauge field modes can be
accomplished. The analysis of the semileptonic data based on this
approach \cite{slcm,slsf} proved to be efficient and successful
\cite{babarprl}. To order $\alpha_s$ it corresponds to a separation
based on the energy $\omega\!=\!|\vec{k}\,|$ of the exchanged gluon:
only those with $\omega\!>\!\mu$ are included into the Feynman
integrals for the Wilson coefficients. While in higher orders the
situation becomes considerably more involved, it can be handled for
all BLM improvements similar to the first order case with only the
relation between $\omega$ and $|\vec{k}\,|$ being modified.

Adopting this separation scheme one computes the usual one-gluon decay
probability, yet with the extra factor
$\theta(|\vec{k}\,|\!-\!\mu)$. A massless gluon in the decay 
necessarily has $|\vec{k}\,|\!<\!\frac{m_b}{2}$, hence taking 
$\mu \!>\!\frac{m_b}{2}$ would eliminate all bremsstrahlung. In what
follows we will always assume $2\mu\!<\!m_b$; in practice, with $\mu$
around $1\GeV$ we have $\frac{\mu}{m_b}\!\simeq\!0.22$.
The integration over the thus restricted phase space is 
straightforward and yields 
\beq
\frac{1}{\Gamma^{\rm pert}}\,
\frac{{\rm d}\Gamma^{\rm pert}(\xi ;\mu)}{{\rm d}\xi } =
C_F\, \frac{\alpha_s}{\pi}\,
f_1(\xi ;\mbox{$\frac{\mu}{m_b}$})\,\theta(1\!-\!\xi )
+ A\,\delta(1\!-\!\xi ) + C_F\, \frac{\alpha_s}{\pi}\,
B_1\left(\mbox{$\frac{\mu}{m_b}$}
\right)\delta^\prime(1\!-\!\xi )\;,
\label{220}
\eeq
where the second term stands for the tree-level two-body
decay modified by virtual corrections. Its strength $A$ is
determined by the condition that the normalized spectrum in
the l.h.s.\ integrated
over the whole domain from $0\!\le\!\xi \!\le\! 1$, is
unity. The last term with $\delta^\prime(1\!-\!\xi )$ can be
viewed as the reflection of the fact that the `kinetic' mass
$m_b(\mu)$ in the presence of perturbative corrections
differs from the energy of the quark at rest (a `rest
energy' mass). The latter determines the perturbative 
end-point, in particular
the photon energy in the two-body decay: 
\beq
B_1(\eta)=\frac{4\,\eta }{3}\! 
+ \!\frac{\eta ^2}{2}\! - \!\frac{1}{2}\left(\ln{\left[\eta 
\!+\!\sqrt{1\!+\!\eta^2}\right]}\!+\!\eta
\left(\sqrt{1+\eta ^2}\!-\!\eta \right)\right)
\label{222}
\eeq
The one-loop expression for the continuous part of the
spectrum is 
\beq
f_1(\xi ;\eta)=\left\{\begin{array}{ll}
\frac{2\,\xi ^3\!-\!3\,\xi^2\!-\!6\,\xi\!}{4(1\!-\!\xi )}
+\frac{\left(\xi^2\!-\!3\right)}{2(1\!-\!\xi )}\ln{(1\!-\!\xi )}
& \qquad \xi \le 1-2\eta \rule[-20pt]{0pt}{30pt}
\\
\frac{(1\!-\!2\,\eta )\left(4\,\xi\,\eta 
^2\!-\!2\,\eta^2\!+\!4\,\xi^2\,\eta \!-\!
6\,\xi \,\eta \!-\!\eta\!+\!2\,\xi\!-\!2\right)}{4\,\eta (1\!-\!\xi )}
+\frac{\left(\xi ^2\!-\!3\right)}{2(1\!-\!\xi )}\ln{(2\,\eta)}
& \qquad \xi >1-2\eta
\end{array}
\right.
\label{224}
\eeq
These allow to readily calculate the required moments of the
perturbative spectrum: 
\begin{small}
\bea
\nonumber
\Phi_0(x;\eta)\msp{-4}&=&\msp{-4} \int_{x}^1  \frac{1}{\Gamma^{\rm pert}}\,
\frac{{\rm d}\Gamma^{\rm pert}(\xi;\mu)}{{\rm d}\xi} \;{\rm d}\xi 
\rule[-20pt]{0pt}{15pt}\\
\msp{-4}&=&\msp{-4} \left\{\begin{array}{ll}
\msp{-4} 1-C_F\,\frac{\alpha_s}{\pi}\,
\left[\frac{x\left(-2\,x^2\!+\!3\,x\!+\!30\right)}{12}
+\frac{\left(-x^2\!-\!2\,x\!+\!10\right)}{4}\ln{(1\!-\!x)}\!+\!
\frac{1}{2}\ln^2{(1\!-\!x)}\right]
& \msp{-2} \mbox{for}\; x\!\le\! 1\!-\!2\eta \rule[-0pt]{0pt}{20pt}
\\
\nonumber
\msp{-4} 1-C_F\,\frac{\alpha_s}{\pi}\left[
\frac{1\!-\!2\,\eta}{12\eta}\left(-6\,\eta\,x^2\!-\!12\,\eta^2\,x\!+
\!6\,\eta\,x\!-\!6\,x\!-\!8\,\eta^3\!+\!2\,\eta^2\!+\!
19\,\eta\!+\!6\right)\right.
&  \rule[-10pt]{0pt}{40pt}
\\
\nonumber 
+\left.\frac{4\,\eta^2\!-\!8\,\eta\!+\!4\ln{(2\,\eta)}
+\!3\!}{4}\ln{(1\!-\!x)}+\frac{-x^2\!-\!2\,x\!-\!4\eta^2
\!+8\,\eta\!+\!7}{4}\ln{(2\,\eta)}-\frac{1}{2}\ln^2{(2\,\eta)} 
\right]& \msp{-2} \mbox{for}\; x\!>\!1\!-\!2\eta
\\
\end{array}\right. \\
\label{230}
\eea
\bea
\nonumber
C_F \mbox{$\frac{\alpha_s}{\pi}$}\, a_1(x;\eta)\msp{-4}&=&\msp{-4}  
\int_{x}^1 \frac{1}{\Gamma^{\rm pert}}\,
\frac{{\rm d}\Gamma^{\rm pert}(\xi;\mu)}{{\rm d}\xi} (1\!-\!\xi)\,{\rm 
d}\xi \rule[-20pt]{0pt}{15pt} \\
\msp{-4}&=&\msp{-4} \left\{
\begin{array}{ll}\msp{-2}C_F\,\frac{\alpha_s}{\pi}\,\left[
\frac{1}{72}\left(48\,\eta^4\!-\!64\,\eta^3\!+\!72\,\eta^2\!-\!72\,\eta
\!-\!9\,x^4\!+\!22\,x^3\!+\!60\,x^2\!-\!96\,x\!+\!23\right)\right.
\msp{-11.5}
& \rule[-10pt]{0pt}{30pt}
\\
\left. \msp{9} +\frac{(1-x)(x^2\!+\!x\!-\!8)}{6}\ln{(1\!-\!x)}
-B_1(\eta)\right]&
\msp{-2} \mbox{for}\; x\!\le\!1\!-\!2\eta \rule[-5pt]{0pt}{10pt}\msp{-4}
\\
\nonumber
\msp{-2} C_F\,\frac{\alpha_s}{\pi}\left[
\frac{(1\!-\!x)(1\!-\!2\,\eta)}{12\,\eta}\left(4\,\eta\,x^2
\!+\!6\,\eta^2\,x\!-\!5\,\eta\,x\!+\!3x\,\!-8\,\eta\!-\!3\right)\right. 
&\rule[15pt]{0pt}{10pt}
\\
\left.\msp{9}\; +\frac{(1-x)(x^2\!+\!x\!-\!8)}{6}\ln{(2\,\eta)}
-B_1(\eta)\right]
& \msp{-2} \mbox{for}\; x\!\ge\!1\!-\!2\eta \msp{-4}
\rule[10pt]{0pt}{10pt}\\
\end{array}\right. \\
\label{231}
\eea
\bea
\nonumber
C_F \mbox{$\frac{\alpha_s}{\pi}$}\,a_2(x;\eta)\msp{-4}&=&\msp{-4}  
\int_{x}^1 \frac{1}{\Gamma^{\rm pert}}\,
\frac{{\rm d}\Gamma^{\rm pert}(\xi;\mu)}{{\rm d}\xi} (1\!-\!\xi)^2\,{\rm 
d}\xi \rule[-15pt]{0pt}{25pt} \\
\msp{-4}&=&\msp{-4} \left\{
\begin{array}{ll}\msp{-2}C_F\,\frac{\alpha_s}{\pi}\,\left[
\frac{1}{1440}\left(768\,\eta^5\!-\!720\,\eta^4
\!+\!640\,\eta^3\!-\!480\,\eta^2\!+\!144\,x^5\!-\!495\,x^4\!-\!340\,x^3\!
\right.\right.\msp{-14}
& \rule[-10pt]{0pt}{20pt}
\\
\msp{9} \left.\left.\;+1650\,x^2\!-\!1020\,x\!+\!61\right)\!+
\!\frac{(1\!-\!x)^2(3\,x^2\!+\!2\,x\!-\!17)}{24}\ln{(1\!-\!x)}\right]&
\msp{-2} \mbox{ for}\; x\!\le\!1\!-\!2\eta \msp{-4}
\\
\nonumber
\msp{-2} C_F\,\frac{\alpha_s}{\pi}\left[
\frac{(1\!-\!x)^2(1\!-\!2\,\eta)}{24\,\eta}\left(6\,\eta\,x^2
\!+\!8\,\eta^2\,x\!-\!8\,\eta\,x\!+\!4\,x\,\!-\!2\,\eta^2
\!-\!7\,\eta\!-\!4\right)\right. 
&\rule[20pt]{0pt}{10pt}
\\
\left.\msp{9} 
\;+\frac{(1\!-\!x)^2(3\,x^2\!+\!2\,x\!-\!17)}{24}\ln{(2\,\eta)}\right]
& \msp{-2} \mbox{ for}\; x\!\ge\!1\!-\!2\eta \msp{-4}
\rule[10pt]{0pt}{10pt}\\
\end{array}\right. \\
\label{232}
\eea
\end{small}

The usually considered BLM corrections can easily be
calculated to arbitrary order with merely technical
modifications. The corresponding technique has been reviewed in
Refs.~\cite{blmvcb,imprec}. First the one-gluon
decay rate must be calculated with a non-vanishing
fictitious gluon mass $m_g$ reflecting the gluon
conversion into the $q\bar{q}$ or $gg$ pairs. This modifies
the cutoff condition: $\theta(|\vec{k}\,|\!-\!\mu)$ is
replaced by $\theta(\omega\!-\!\mu)=
\theta(\vec{k}^{\,2}\!+\!m_g^2\!-\!\mu^2)$. The integrals
are still easily taken. All the functions additionally depend now on the 
new dimensionless parameter 
$\lambda\!\equiv\!\frac{m_g^2}{m_b^2}$  (note that in Ref.~\cite{imprec}
this ratio was denoted by $\lambda^2$). Generalizing Eq.~(\ref{220}) 
we then set 
\beq
\frac{1}{\Gamma^{\rm pert}}\,
\frac{{\rm d}\Gamma^{\rm pert}(\xi,m_g;\mu)}{{\rm d}\xi } =
C_F\, \frac{\alpha_s}{\pi}\,
f_1(\xi,\lambda ;\eta)\,\theta(1\!-\!\xi\!-\!\lambda )
+ A\,\delta(1\!-\!\xi ) + C_F\, \frac{\alpha_s}{\pi}\,
B_1\left(\eta;\lambda \right)\delta^\prime(1\!-\!\xi )
\label{236}
\eeq
(recall that $\eta\!<\!\frac{1}{2}$ is assumed; otherwise the kinematic
constraints are modified), with {\small
\bea
\nonumber
B_1\left(\mbox{$\eta $};\lambda\right) \msp{-4}&=&\msp{-4}
\frac{4}{3}\sqrt{\eta^2\!-\!\lambda}\left(1\!-\!
\mbox{$\frac{\lambda}{4\,\eta 
^2}$}\right)\!-\!\frac{\pi\sqrt{\lambda}}{2}\!+\!
\sqrt{\lambda}\arcsin{\left(
\mbox{$\frac{\sqrt{\lambda}}{\eta}$}\right)}\!+\!
\frac{(\eta ^2\!-\!\lambda)^{\frac{3}{2}}}{2\,\eta }\\
\nonumber
\msp{-4}&-&\msp{-4}
\frac{1}{4}\left(2\!-\!\lambda\right)
\sqrt{\lambda\left(4\!-\!\lambda\right)}
\left(\arctan{\left[\frac{(\lambda\!-\!2)
\sqrt{\eta^2\!-\!\lambda}}{\sqrt{\lambda\left(4\!-\!\lambda\right)
\left(\eta^2\!-\!\lambda+1\right)}}\right]}\!-\!
\arctan{\left[\frac{\lambda\sqrt{\eta ^2\!-\!\lambda}}{\eta
\sqrt{\lambda\left(4\!-\!\lambda\right)}}\right]}\right)\\
\nonumber
&-&\msp{-3}\frac{1}{8}\lambda\left(4\!-\!\lambda\right)\left(
2\,\ln{[\eta \!+\!\sqrt{\eta 
^2\!-\!\lambda}]}\!-\!\ln{\lambda}\right)\!-\!
\frac{\lambda^2\!-\!4\lambda\!+\!2}{4}
\ln{(\sqrt{\eta ^2\!-\!\lambda\!+\!1}\!+\!\sqrt{\eta ^2\!-\!\lambda})}\\
&-&\msp{-3}\frac{\sqrt{\eta ^2\!-\!\lambda}}{2}
\left(\sqrt{\eta ^2\!-\!\lambda\!+\!1}\!-\!\eta \right).
\qquad \label{240}
\eea
}

To calculate the BLM corrections, we must integrate over the gluon
mass.
The upper end of the $\mu$-independent lower region of the spectrum now
rises depending on $\lambda$: the perturbative spectrum 
is independent of $\mu$ 
if $\lambda\!>2\,\eta \,(1\!-\!\xi)\!-\!(1\!-\!\xi)^2$:
\begin{small} 
\bea
\nonumber
f_1\left(\xi;\mbox{$\eta$};
\mbox{$\lambda$}\right)\msp{-5}&=&\msp{-5}
\frac{\xi\,(\xi\!+\!\lambda\!-\!1)}{4(1\!-\!\xi)^3(\xi^2\!-\!2\,\xi\!+
\!\xi\,\lambda\!+\!1)}\left(2\,\xi^3\lambda^2\!-\!3\,\xi^2\lambda^2\!+
\!2\,\xi\lambda^2\!+\!4\,\xi^4\,\lambda\!-\!14\,\xi^3\lambda\!+\!
17\,\xi^2\,\lambda\!-\!11\,\xi\lambda\!\right.\rule[-10pt]{0pt}{30pt}\\
\nonumber
&\msp{-5}\msp{-5}&
+\left. 4\,\lambda\!+\!2\,\xi^5\!-\!
9\,\xi^4\!+\!9\,\xi^3\!+\!7\,\xi^2\!-\!15\,\xi\!+\!6\right)\!+\!
\frac{2\,\xi\,\lambda\!-\!2\,\lambda\!+\!
\xi^2\!-\!3}{2(1\!-\!\xi)}\ln{\left(1\!-\!\xi\!+\!
\frac{\xi\,\lambda}{1\!-\!\xi}\right)} \\ 
\nonumber
& & \msp{98.5} 
\mbox{at}\; \lambda\!\ge\!2\eta(1\!\!-\!\!\xi)\!\!-\!\!
(1\!\!-\!\!\xi)^2\rule[-20pt]{0pt}{30pt}\vspace*{-50pt}\\
& & \vspace*{-50pt}
\label{241}\\
\nonumber
&= & \msp{-4}\frac{(2\,\eta\!-\!\lambda\!-\!1)}{4(1\!-\!\xi)^2
(2\,\eta\!-\!\lambda)}\left(-2\,\xi^2\,\lambda^2\!+\!\xi\,
\lambda^2\!-\!\lambda^2\!-\!4\,\xi^3\,\lambda\!+\!12\,\xi^2\,
\lambda\!+\!4\,\eta\,\xi\,\lambda\!-\!9\,\xi\,\lambda\!+\!\lambda\!\right.
\rule[-10pt]{0pt}{30pt}\\
\nonumber
& & \msp{-9}\left.+8\,\eta\,\xi^3\!+\!8\,\eta^2\,\xi^2\!-\!
20\,\eta\,\xi^2\!+\!4\,\xi^2\!-\!12\,\eta^2\,\xi\!+\!10\,\eta\,\xi\!-\!
8\,\xi\!+\!4\,\eta^2\!+\!2\,\eta\!+\!4\right) \rule[-10pt]{0pt}{30pt}\\
\nonumber
& & \msp{-9} +\frac{2\,\xi\,\lambda\!-\!2\,\lambda\!+\!
\xi^2\!-\!3}{2(1\!-\!\xi)}  \ln{\left(2\,\eta-\lambda\right)} 
\msp{50}\mbox{{ at $\lambda\!\le 
\!2\eta(1\!\!-\!\!\xi)\!\!-\!\!(1\!\!-\!\!\xi)^2$}} 
\eea
\end{small}\vspace*{-10pt}

\noindent
The above $\lambda$-dependent functions replace the usual
one-loop ones of Eqs.~(\ref{220})-(\ref{224}).
Using Eqs.~(\ref{240}), (\ref{241})
it is not difficult to obtain the BLM corrections both to the
spectrum and to its moments. For instance, for the continuous part 
of the spectrum,
at $\xi\!\le\!1\!-\!2\eta$ we integrate the upper of the two expressions
in Eq.~(\ref{241}) over $\lambda$ directly from $\lambda\!=\!0$ to
$\lambda\!=\!1\!-\!\xi$. At $\xi\!>\!1\!-\!2\eta$ the integral from
$\lambda\!=\!0$ to $\lambda\!=\!(1\!-\!\xi)(\xi\!-\!1\!+\!2\eta)$ uses
the lower expression, while the integrand in the integral from
$\lambda\!=\!(1\!-\!\xi)(\xi\!-\!1\!+\!2\eta)$ to
$\lambda\!=\!1-\!\xi$ is the $\eta$-independent 
upper expression. The subtraction piece
employing $\lambda\!=\!0$, of course is integrated over $\lambda$ from $0$
to $\infty$, \,cf.\ Eq.~(\ref{251}).

Representing the spectrum as
\bea
\nonumber
\frac{1}{\Gamma^{\rm pert}}\,
\frac{{\rm d}\Gamma^{\rm pert}(\xi;\mu)}{{\rm d}\xi} &\msp{-5}=\msp{-5}&
C_F\, \frac{\alpha_s}{\pi}\, \left[f_1(\xi;\mbox{$\frac{\mu}{m_b}$})+
\beta_0\mbox{$\frac{\alpha_s}{\pi}$}\,f_2(\xi;\mbox{$\frac{\mu}{m_b}$})
+ ...\;\right]
\,\theta(1\!-\!\xi)
+ A\,\delta(1\!-\!\xi) + \\
&& \msp{10}
C_F\, \frac{\alpha_s}{\pi}\,
\left[B_1\left(\mbox{$\frac{\mu}{m_b}$} \right) +
\beta_0\mbox{$\frac{\alpha_s}{\pi}$}
B_2\left(\mbox{$\frac{\mu}{m_b}$}\right)+ ...\;
\right]\,\delta^\prime(1\!-\!\xi)\;,
\qquad \label{260}
\eea
$f_1$ and $B_1$ are given in Eqs.~(\ref{224}) and (\ref{222}),
respectively, while for the second-order BLM parts we obtain:
\begin{small}
\bea
f_2(\xi;\eta)\msp{-4}&=&\msp{-4} 
\frac{38\,\xi^3\!-\!93\,\xi^2\!+\!6\,\xi\!-\!36}{96(1\!-\!\xi)}-
\frac{6\,\xi^4\!-\!31\,\xi^3\!+\!24\,\xi^2\!-\!30\,\xi\!+\!18}
{48\,\xi(1\!-\!\xi)}\ln{(1\!-\!\xi)}
 \rule[-15pt]{0pt}{20pt} \label{250}
\\
\nonumber
& &
\nonumber
-\frac{3(\xi^2\!-\!3)}{16(1\!-\!\xi)}
\left(\ln^2{(1\!-\!\xi)}\!+\!\frac{2}{3}{\rm 
Li}_2(\xi)\right) 
\msp{67.9} \xi\!\le\! 1\!-\!2\eta \rule[-30pt]{0pt}{40pt}\\
\nonumber
\msp{-4}&=&\msp{-4}\frac{6\,\eta\,\xi^3\!-\!(128\,\eta^2\!-\!16\,\eta)\xi^2
\!-\!(128\,\eta^3\!-\!244\,\eta^2\!+\!52\,\eta\!-\!20)\xi
+52\,\eta^3\!+\!12\,\eta^2\!-\!15\,\eta\,-\,20}{96\,\eta\,(1\!-\!\xi)}\\
\nonumber
&& \msp{-5}+\frac{(8\,\eta^2\!-\!4\,\eta)\,\xi^2\!+\!(8\,\eta^3\!-\!16\,\eta^2
\!+\!10\,\eta-2)\,\xi\!-\!4\,\eta^3\!-\!3\,\eta\!+\!2}
{16\,\eta\,(1\!-\!\xi)}\left(\ln{(1\!-\!\xi)}
\!+\!\ln{(2\,\eta\!+\!\xi\!-\!1)}\right)\rule[-10pt]{0pt}{20pt}\\
\nonumber
&& \msp{-5}+\frac{2\,\xi^3\!-\!3\,\xi^2\!-\!6\,\xi}{16\,(1\,-\,\xi)}
\ln{(2\,\eta\!+\!\xi\!-\!1)}+\frac{10\,\eta\,\xi^2
\!-\!2\,(12\,\eta^2\!+\!3\,\eta\!+\!3)\,\xi\!+\!24\,\eta^2
\!-\!24\,\eta\!+\!6}{48\,\eta\,(1\!-\!\xi)}\ln{(2\,\eta)}
\rule[-10pt]{0pt}{20pt}\\
\nonumber
&& \msp{-5}+\frac{4\,\xi^3\,\eta\!+\!2(4\,\eta^2\!-\!6\,\eta\!+\!1)\xi^2
\!-\!2(4\,\eta^2\!-\!9\,\eta\!+\!1)\,\xi\!-\!6\,\eta}
{16\,\eta\,\xi(1\!-\!\xi)}\ln{(\xi^2\!+\!2\,\eta\,\xi\!-\!2\,\xi\!+\!1)}
\rule[-10pt]{0pt}{20pt}\\
\nonumber
&& \msp{-5}+\frac{\left(\xi^2\!-\!3\right)}{8(1\!-\!\xi)}
\left\{\vphantom{\frac{\xi\,(2\,\eta\!+\!\xi\!-\!1)}{\xi\!-\!1}} 
\ln{(1\!-\!\xi)}\ln{(2\,\eta\!+\!\xi\!-\!1)}\!-\!\ln{(1\!-\!\xi)}
\ln{(2\,\eta)}\!-\!\ln{(2\,\eta)}\ln{(2\,\eta\!+\!\xi\!-\!1)} 
\right.\rule[-10pt]{0pt}{20pt}\msp{-5}\\
\nonumber
&& \left.\msp{-1}-\frac{1}{2}\ln^2{\!(1\!-\!\xi)}\!-\!{\rm 
Li}_2(\xi)\!+\!{\rm Li}_2\!
\left(\!\frac{(1\!-\!\xi)(2\,\eta\!+\!\xi\!-\!1)}{2\,\eta}\right)\!-\!{\rm 
Li}_2\!\left(\!\frac{\xi\,(2\,\eta\!+\!\xi\!-\!1)}{\xi\!-\!1}\right)\right\}
\msp{6.5}  \xi\!>\! 1\!-\!2\eta \msp{-5}
\eea
\end{small}

The BLM correction to the coefficient of the $\delta^\prime(1\!-\!\xi)$ can
be obtained by numerically integrating over $\lambda$
\beq
B_2\left(\mbox{$\eta$}\right)=\frac{1}{4}\int_0^{\infty}
\frac{d\lambda}{\lambda}\left(\frac{B_1\left(\mbox{$\eta$};
\mbox{$0$}\right)}{1+e^{-5/3}\lambda}\!-\!B_1\left(\mbox{$\eta$};
\mbox{$\lambda$}\right)\right)
\label{251}
\eeq

The complete $\mu$-dependent analytic expressions for the BLM corrections
to the
moments contain polylog functions at worst, but are somewhat
cumbersome, and we do not quote them here. At $\mu\!=\!0$
the order-$\alpha_s$ spectrum and
its second-order BLM correction reproduce the standard
expressions \cite{llmw}
\bea
\label{266}
f_1(\xi;0)
&\msp{-5}=\msp{-5}&\frac{2\,\xi^3\!-\!3\,\xi^2\!-\!6\,\xi\!}{4(1\!-\!\xi)}
+\frac{\left(\xi^2\!-\!3\right)}{2(1\!-\!\xi)}\ln{(1\!-\!\xi)} \\
\nonumber
f_2(\xi;0)
&\msp{-5}=\msp{-5}&\frac{38\,\xi^3\!-\!93\,\xi^2\!+\!6\,\xi\!-\!36}{96(1\!-\!
\xi)}-\frac{6\,\xi^4\!-\!31\,\xi^3\!+\!24\,\xi^2\!-\!30\,\xi\!+
\!18}{48\,\xi(1\!-\!\xi)}\ln{(1\!-\!\xi)}\\&\msp{-5}-
\msp{-5}&\frac{3(\xi^2\!-\!3)}{16(1\!-\!\xi)}\left(\ln^2{(1\!-\!\xi)}\!+\!
\frac{2}{3}{\rm Li}_2(\xi)\right) \;.
\label{267}
\eea
Their moments, $\,a_k(x;0)\,$ and $\,b_k(x;0)$,  respectively are given by
\begin{small}
\bea
\nonumber
\int_0^x f_1(\xi;0) \;{\rm d}\xi
&\msp{-5}=\msp{-5}&\frac{x\left(-2\,x^2\!+\!3\,x\!+\!30\right)}{12}
+\frac{\left(-x^2\!-\!2\,x\!+\!10\right)}{4}\ln{(1\!-\!x)}\!+
\!\frac{1}{2}\ln^2{(1\!-\!x)} 
\\
\nonumber
\int_0^x f_2(\xi;0)\;{\rm d}\xi
&\msp{-5}=\msp{-5}&\frac{-14\,x^3\!+\!41\,x^2\!+\!120\,x\!-\!48\,\zeta (3)
}{96}\!+\!\frac{1}{4}{\rm Li}_2\left(x\right) \ln{(1\!-\!x)} \\
\nonumber
&\msp{-5}+\msp{-5}&\frac{4\,x^3\!-\!31\,x^2\!-\!44\,x\!-\!8\,\pi^2
\!+\!156}{96}\ln{(1\!-\!x)}\!+\!
\frac{x^2\!+\!2\,x\!+\!6}{16}{\rm Li}_2\left(x\right)\\
\nonumber
&\msp{-5}+\msp{-5}&\frac{9\,x^2\!+\!18\,x\!-\!40\!+\!
24\ln{(x)}}{96}\ln^2{(1\!-\!x)}\!-\!\frac{1}{8}\ln^3{(1\!-\!x)}\!+\!
\frac{1}{2}{\rm Li}_3\left(1\!-\!x\right)\\
\nonumber
\int_x^1 f_1(\xi;0) \;(1\!-\!\xi){\rm d}\xi
&\msp{-5}=\msp{-5}&\frac{1}{72}\left(-\!9\,x^4\!+\!22\,x^3\!+\!60\,x^2\!-\!
96\,x\!+\!23\right)\!+\!
\frac{(1\!-\!x)(x^2\!+\!x\!-\!8)}{6}\ln{(1\!-\!x)}\\
\nonumber
\int_x^1 f_2(\xi;0)\;(1\!-\!\xi){\rm d}\xi
&\msp{-5}=\msp{-5}&\frac{408\,\pi^2\!-\!369\,x^4\!+\!1360\,x^3\!-\!
66\,x^2\!+\!1380\,x\!-\!2305}{3456}\!+\!
\frac{x^3\!-\!9\,x\!-\!9}{24}{\rm Li}_2\left(x\right) \\
\nonumber
&\msp{-5}-\msp{-5}&(1\!-\!x)\left(\frac{9\,x^3\!-\!61x\,^2\!-\!
7\,x\!-\!7}{288}\ln{(1\!-\!x)}\!+\!
\frac{x^2\!+\!x\!-\!8}{16}\ln^2{(1\!-\!x)}\right)\\
\nonumber
\int_x^1 f_1(\xi;0) \;(1\!-\!\xi)^2{\rm d}\xi
&\msp{-5}=\msp{-5}&\frac{1}{1440}\left(144\,x^5\!-\!495\,x^4\!-\!340\,x^3+
1650\,x^2\!-\!1020\,x\!+\!61\right)\\
\nonumber
&\msp{-5}+\msp{-5}&\frac{(1\!-\!x)^2(3\,x^2\!+\!2\,x\!-\!17)}{24}
\ln{(1\!-\!x)} 
\\
\nonumber
\int_x^1 f_2(\xi;0)\;(1\!-\!\xi)^2{\rm d}\xi
&\msp{-5}=\msp{-5}&\frac{7950\,\pi^2\!+\!7272\,x^5\!-\!33435x\,^4\!+\!
35320\,x^3\!-\!25995\,x^2\!+\!91110\,x\!-\!74272}{86400}\\
\nonumber
&\msp{-5}+\msp{-5}& 
\frac{-3x\,^4\!+\!4\,x^3\!+\!18\,x^2\!-\!36\,x\!-\!36}{96}
{\rm
Li}_2\left(x\right)\!-\!\frac{(1\!-\!x)^2(3\,x^2\!+\!2\,x\!-\!17)}{48}
\ln^2{(1\!-\!x)}\\
&\msp{-5}+\msp{-5}&\frac{(1\!-\!x)(72\,x^4\!-\!528\,x^3\!+\!562\,x^2\!-\!
473\,x\!+\!1957)}{2880} \ln{(1\!-\!x)}
\label{269}
\eea
\end{small}

Finally, we provide simple numerical expressions for the first and 
second moments of the second order BLM correction with $\mu/m_b$ fixed 
at the value $1/4.6$: 
\begin{small}
$$
b_1(x;\mbox{$\frac{1}{4.6}$})\!=\!\msp{140}
$$
\vspace*{-20pt}
\beq
-B_2(\mbox{$\frac{1}{4.6}$})+
\int_x^1 \!\! f_2(\xi;\mbox{$\frac{1}{4.6}$})\;(1\!-\!\xi){\rm d}\xi
=\left\{\begin{array}{ll}\!\!-0.032\!-\!0.096\,y\!-
\!0.67\,y^2\!+\!1.6\,y^3\!-\!
3.2\,y^4 & 
\mbox{for } y\ge 0 \rule[-10pt]{0pt}{10pt}\\
\nonumber
\!\! -0.032\!-\!0.11\,y\!-\!0.33\,y^2\!-\!0.47\,y^3\!-\!0.26
\,y^4 & 
\mbox{for } y \le 0 
\end{array}
\right.
\label{282}
\eeq
{\normalsize 
$$
\msp{20}y\equiv x-(1-2/4.6)
$$}
\beq
b_2(x;\mbox{$\frac{1}{4.6}$})\!=\! 
\int_x^1 \!\! f_2(\xi;\mbox{$\frac{1}{4.6}$})\;(1\!-\!\xi)^2{\rm d}\xi
=\left\{\begin{array}{ll}
\!\!0.0225\!-\!0.051\,y\!-\!0.066\,y^2\!+\!0.040 \,y^3\!+\!
0.24\,y^4 & 
\mbox{for } y\ge 0 \rule[-10pt]{0pt}{10pt}\\
\nonumber
\!\! 0.0225\!-\!0.050\,y\!-\!0.12\,y^2\!-\!0.12\,y^3\!-\!0.04
\,y^4 & 
\mbox{for } y \le 0 
\end{array}
\right.
\label{283}
\eeq
\end{small}
\vspace*{-12pt} 

\noindent
They are sufficiently accurate in the whole relevant domain of $x$
between $0.65$ and $0.93$. 
The evaluation for different values of $m_b$ can easily be recovered from
the fact that the above integrals are functions of the ratio
$\mu/m_b$, and at fixed $m_b$ the $\mu$-dependence is well described by 
Eqs.~(\ref{30}).

\subsection{\boldmath Non-$O_7$ perturbative corrections}

The non-$O_7$ spectrum was given in Ref.~\cite{llmw}. We have interpolated 
the moments of these contributions for 
$\frac{m_c}{m_b}\!=\!\frac{1.18}{4.61}$:
\bea
\nonumber
100\,\hat a_1(x) \msp{-4}&=&\msp{-4} \mbox{$\frac{C_2^2}{C_7^2}$}
[0.053\!-\!0.81 \,(x\!-\!0.87)+2.93\,(x\!-\!0.87)^2] +
\mbox{$\frac{C_2}{C_7}$}
[-0.026\!+\!0.25 \,(x\!-\!0.87)\\ 
&&\;\; +0.29\,(x\!-\!0.87)^2]
\nonumber
+
\mbox{$\frac{C_8}{C_7}$}
[0.56\!-\!7.5 \,(x\!-\!0.87)+21\,(x\!-\!0.87)^2]\\
\nonumber
100\,\hat b_1(x) \msp{-4}&=&\msp{-4} \mbox{$\frac{C_2^2}{C_7^2}$}
[0.064\!-\!0.86 \,(x\!-\!0.87)+2.25\,(x\!-\!0.87)^2]+
\mbox{$\frac{C_2}{C_7}$}
[0.0049\!-\!0.25 \,(x\!-\!0.87)\\
\nonumber
&&\;\; +2.18\,(x\!-\!0.87)^2]
+
\mbox{$\frac{C_8}{C_7}$}
[0.72\!-\!8.8 \,(x\!-\!0.87)+18\,(x\!-\!0.87)^2]\\
\nonumber
100\,\hat a_2(x) \msp{-4}&=&\msp{-4} \mbox{$\frac{C_2^2}{C_7^2}$}
[0.0023\!-\!0.055 \,(x\!-\!0.87)+0.38\,(x\!-\!0.87)^2]+
\mbox{$\frac{C_2}{C_7}$}
[-0.0010\!+\!0.015 \,(x\!-\!0.87)\\
\nonumber
&&\;\; -0.0413\,(x\!-\!0.87)^2]
+
\mbox{$\frac{C_8}{C_7}$}
[0.047\!-\!1.1 \,(x\!-\!0.87)+6.5\,(x\!-\!0.87)^2]\\
\nonumber
100\,\hat b_2(x) \msp{-4}&=&\msp{-4} \mbox{$\frac{C_2^2}{C_7^2}$}
[0.005\!-\!0.12 \,(x\!-\!0.87)+0.71\,(x\!-\!0.87)^2]+
\mbox{$\frac{C_2}{C_7}$}
[0.0007\!-\!0.042 \,(x\!-\!0.87)\\
&&\;\; +0.42\,(x\!-\!0.87)^2]
+
\mbox{$\frac{C_8}{C_7}$}
[0.06\!-\!1.2 \,(x\!-\!0.87)]+6.75\,(x\!-\!0.87)^2 . \;
\label{302}
\eea
Since their effect is small and well below a host of the corrections
left out, closer attention would seem superfluous.

\subsection{Tables for biases}
\label{biastables}

Here we tabulate the values for biases obtained with the purely
nonperturbative distribution functions, as referred to in Sects.~3.1
and 3.3. To make the tables compact, we present biases in the scaling
form (\ref{54}), evaluated near the expected value
$r\!=\!0.75$. 
The two primary entries are the averages of $f_s(q)$ (first moment) or
of $g_s(q)$ (second moment) obtained with the two distribution functions
$F_1$ and $F_2$ in Eqs.~(\ref{52}). 
\begin{table}[h]
\vspace*{-4mm}
\caption{The estimated values for the bias corrections 
around $r\!=\!0.75$}\vspace*{-2.5mm}
\begin{center}
\begin{tabular}{|l|l|l|l|l|l|l|}
\hline
$q$ & $f_s$ & $g_s$ & $D_f$ & $D_g$ & $F_{12}$ & $G_{12}$\\
\hline
\hline
$0$ & $ 0.413  $ & $0.800  $ & $ -0.0337 $ & $ 0.0920 $ & $ -0.0151 
$ & $0.0209 $ \\
$0.5$ & $ 0.208  $ & $ 0.577 $ & $ 0.00453 $ & $ 0.107 $ & $-0.00312  
$ & $0.0325 $ \\
$0.75$ & $ 0.137  $ & $ 0.452 $ & $ 0.0141 $ & $ 0.112 $ & $ 0.00222 
$ & $ 0.0389 $ \\
$1.0$ & $ 0.0854  $ & $ 0.333  $ & $ 0.0179 $ & $ 0.112 $ & $0.00571 
$ & $0.0440  $ \\
$1.25$ & $ 0.0507  $ & $ 0.231 $ & $ 0.0175 $ & $ 0.105 $ & $ 0.00712 
$ & $ 0.0460 $ \\
$1.5$ & $ 0.0287  $ & $ 0.151 $ & $ 0.0146  $ & $ 0.0915 $ & $ 0.00684
$ & $ 0.0436 $ \\
$1.75$ & $ 0.0156  $ & $0.0938  $ & $ 0.0110  $ & $ 0.0743 $ & $
0.00561 $ & $ 0.0375 $ \\
$2.0$ & $ 0.00829  $ & $ 0.0560 $ & $ 0.00769 $ & $ 0.0566  $ & $ 
0.00408 $ & $ 0.0294 $ \\
$2.5$ & $ 0.00225  $ & $ 0.0187 $ & $ 0.00329 $ & $ 0.0288 $ & $ 
0.00168 $ & $ 0.0143 $ \\
$3.0$ & $ 0.000606  $ & $ 0.00597 $ & $ 0.00129 $ & $ 0.0132 $ & $ 
0.000553 $ & $ 0.0055 $ \\
\hline
\end{tabular}
\vspace*{-4mm}
\end{center}
\end{table}
Their interpolating expressions
are given in Eqs.~(\ref{64}). To show and/or to correct for the 
residual dependence on $r$ the third and the fourth entries, $D_f(q)$
and  $D_g(q)$, respectively, linearly approximate this dependence of
$f$ and $g$:
\beq
f_s(q) \longrightarrow f_s(q) + D_f(q) \, (r\!-\!0.75), \qquad 
g_s(q) \longrightarrow g_s(q) + D_g(q) \, (r\!-\!0.75)
\label{346}
\eeq
(these dependences are approximate and were derived by varying
$r$ by $\pm 0.25$). Finally, the differences between the two 
ans\"{a}tze are illustrated by the last two columns, for $\tilde\delta m_b$
and $\tilde\delta \mu_\pi^2$ respectively:
\beq
f_s^{1,2}(q) = f_s(q) \pm F_{12}(q)\, \qquad 
g_s^{1,2}(q) = f_s(q) \pm G_{12}(q)\, \qquad 
\label{348}
\eeq
where the upper and lower signs refer to the biases for $F_1$ and
$F_2$, respectively. 

\begin{table}[h]
\vspace*{-4mm}
\caption{The similar bias corrections for $r$
around $1.0$ } \vspace*{-2.5mm}
\begin{center}
\begin{tabular}{|l|l|l|l|l|l|l|}
\hline
$q$ & $f_s$ & $g_s$ & $D_f$ & $D_g$ & $F_{12}$ & $G_{12}$\\
\hline
\hline
$0$ & $ 0.405  $ & $0.823  $ & $ -0.0311 $ & $ 0.0760 $ & $ -0.0165 
$ & $0.0184 $ \\
$0.5$ & $ 0.209  $ & $ 0.603 $ & $ 0.00290 $ & $ 0.0932 $ & $ -0.00454
$ & $0.0321 $ \\
$0.75$ & $ 0.140  $ & $ 0.480 $ & $ 0.0116 $ & $ 0.0988 $ & $ 0.00120 
$ & $ 0.0399 $ \\
$1.0$ & $ 0.0899  $ & $ 0.361  $ & $ 0.0153 $ & $ 0.0999 $ & $0.00522 
$ & $0.0465  $ \\
$1.25$ & $ 0.0551  $ & $ 0.257 $ & $ 0.0153 $ & $ 0.0951 $ & $ 0.00717
$ & $ 0.0501 $ \\
$1.5$ & $ 0.0324  $ & $ 0.174 $ & $ 0.0131 $ & $ 0.0845 $ & $ 0.00733 
$ & $ 0.0492 $ \\
$1.75$ & $ 0.0184  $ & $ 0.112 $ & $ 0.0101 $ & $ 0.0701 $ & $ 0.00633
$ & $ 0.0440 $ \\
$2.0$ & $ 0.0102  $ & $ 0.0702 $ & $ 0.00729 $ & $ 0.0548 $ & $ 
0.00487 $ & $ 0.0361 $ \\
$2.5$ & $ 0.00307  $ & $ 0.0259 $ & $ 0.00332 $ & $ 0.0296 $ & $ 
0.00225  $ & $ 0.0195 $ \\
$3.0$ & $ 0.000930  $ & $ 0.00927 $ & $ 0.00141 $ & $ 0.0146 $ & $ 
0.000840 $ & $ 0.0084 $ \\
\hline
\end{tabular}
\vspace*{-4mm}
\end{center}
\end{table}
We also give the similar table around a different value of
$r=\widehat{\mu_\pi^2}/\widehat\Lambda^2=1.0$; in this case
$(r\!-\!0.75)$ in Eqs.~(\ref{346}) must be replaced by $(r\!-\!1)$.

\vspace*{5mm}


\end{document}